\documentclass[10pt,journal,compsoc]{IEEEtran}
\usepackage[utf8]{inputenc}
\usepackage[T1]{fontenc}

\usepackage{makecell}

\usepackage{mwe}
\usepackage{rotating}
\usepackage{multicol}
\usepackage[dvipsnames]{xcolor}
\usepackage{tikz-network}
\usepackage{tikz}
\usetikzlibrary{tikzmark,calc,positioning,arrows,shapes,decorations.pathreplacing}
\tikzset{every picture/.style={remember picture}}
\definecolor{mygray}{gray}{0.67}
\definecolor{mylightgray}{gray}{0.87}

\newcommand{\citewb}[1]{\colorbox{white}{\!\!\cite{#1}}}
\newcommand{\citegb}[1]{\colorbox{mygray}{\!\!\cite{#1}}}
\newcommand{\citelgb}[1]{\colorbox{mylightgray}{\!\!\cite{#1}}}

\newcommand*{\myfont}{\fontfamily{phv}\selectfont} %

\newcommand{\slopeSymbol}       {\ell}
\newcommand{\saturationSymbol}  {s}

\newcommand{\methodinitials}   {gLRIE\xspace}
\newcommand{\methodname}       {Generalized Largest Reduction in Infectious Edges\xspace}

\usepackage{times}
\usepackage{placeins}
\usepackage[export]{adjustbox}
\usepackage{ifxetex}
\usepackage{flafter}
\usepackage[utf8]{inputenc}
\usepackage{xcolor}
\usepackage[space]{grffile}
\usepackage{mathtools}
\usepackage{comment}
\usepackage{array,multirow}

\usepackage{lineno}

\usepackage{graphics}
\usepackage{graphicx}

\usepackage{algorithm}
\usepackage{algpseudocode}
\usepackage{enumitem}
\usepackage{amsthm}

\newtheorem{definition}{Definition}
\newtheorem{remark}{Remark}

\usepackage{xcolor}
\usepackage[hidelinks]{hyperref}
\hypersetup{
    colorlinks,
    linkcolor={red!80!black},
    citecolor={blue!50!black},
    urlcolor={blue!80!black}
}
\usepackage{url}
\usepackage{amsmath}

\usepackage{amsfonts}
\usepackage{amssymb}
\usepackage{dsfont}
\usepackage{upgreek}
\usepackage{bbm}			%
\usepackage{bm}

\usepackage{graphicx}
\usepackage{epstopdf}
\graphicspath{ {./} } %
\usepackage{tikz}
\usepackage{float}

\usepackage{subfigure}
\usepackage{booktabs}

\usepackage{xspace}
\usepackage{xifthen}

\usepackage[scaled=.8]{beramono}

\newcommand{\overbar}[1] {\mkern 1.5mu\overline{\mkern-3.5mu#1\mkern-1.0mu}\mkern 1.5mu}

\newcommand{\Sec}[1]		{Sec.\,\ref{#1}}
\newcommand{\Fig}[1]		{Fig.\,\ref{#1}}

\newcommand{\Eq}[1]			{Eq.\,\ref{#1}}
\newcommand{\Tab}[1]		{Tab.\,\ref{#1}}
\newcommand{\Alg}[1]		{Alg.\,\ref{#1}}
\newcommand{\Model}[1]		{Model\,\ref{#1}}
\newcommand{\Appendix}[1]		{Appendix\,\ref{#1}}

\newcommand{\Remark}[1]{Remark~\ref{#1}}
\newcommand{\ie}   			{i.e.\@\xspace}
\newcommand{\eg}   			{e.g.\@\xspace}
\newcommand{\etc}   		{etc.\xspace}
\newcommand{\wrt}   		{w.r.t.\@\xspace}

\newcommand{\Erdos}   	{Erd\"os-R\'enyi}

\newcommand{\ind}       {\mathds{1}}%
\newcommand{\Ind}[1]    {\ind{\{#1\}}}
\newcommand{\notX}[1][] {\overbar{X} \ifthenelse{\isempty{#1}}{}{_{\!#1}}} %
\newcommand{\bigO} 		  {\mathcal{O}} %

\newcommand{\Exp}       {\mathds{E}}%
\newcommand{\real}      {\mathds{R}}
\newcommand{\op}[1]     {\,{#1}\,}

\newcommand{\inlinetitle}[2]  {\smallskip\noindent\textbf{\emph{#1}{#2}}}

\newcommand{\beforecaptvskip} {\vskip -0.07in}

\newcommand{\padded}[1] {\,#1\,}
\newcommand{\void}[1] {\padded{\cdot}}

\newcommand{\budget}  {b}%

\newcommand{\FFun} {\mathcal{F}}
\newcommand{\DeltaF}[2] {\Delta\FFun_{#1}^{-#2}} %
\newcommand{\InfFun} {\mathcal{I}}
\newcommand{\HealFun} {\mathcal{H}}
\newcommand{\DeltaI}[2] {\Delta\InfFun_{#1}^{-#2}} %
\newcommand{\DeltaH}[2] {\Delta\HealFun_{#1}^{-#2}}%

\newcommand{\xoverbrace}[2][\vphantom{\dfrac{A^2}{A}}]{\overbrace{#1#2}}

\usepackage[normalem]{ulem}

\newcounter{marginNoteCounter}

\numberwithin{equation}{section}

\newcommand{\International}     {Intern.\xspace}
\newcommand{\Conference}        {Conf.\xspace}

\newcommand{\Transactions}      {Trans.\xspace}
\newcommand{\Proceedings}       {Proc. of the \xspace}
\newcommand{\Journal}           {Journal\xspace}

\DeclareMathAlphabet{\altmathcal}{OMS}{cmsy}{m}{n}
\renewcommand{\mathcal}[1]  {\altmathcal{#1}}

\makeatletter
\DeclareRobustCommand*{\escapeus}[1]{%
    \begingroup\@activeus\scantokens{#1\endinput}\endgroup}
\begingroup\lccode`\~=`\_\relax
    \lowercase{\endgroup\def\@activeus{\catcode`\_=\active \let~\_}}
\makeatother

\newcommand{\mySqBullet}		{\raisebox{0.25em}{{\scriptsize$_\blacksquare$}}}

\usepackage{scalerel}
\usepackage{tikz}
\usetikzlibrary{svg.path}

\definecolor{orcidlogocol}{HTML}{A6CE39}
\tikzset{
  orcidlogo/.pic={
    \fill[orcidlogocol] svg{M256,128c0,70.7-57.3,128-128,128C57.3,256,0,198.7,0,128C0,57.3,57.3,0,128,0C198.7,0,256,57.3,256,128z};
    \fill[white] svg{M86.3,186.2H70.9V79.1h15.4v48.4V186.2z}
                 svg{M108.9,79.1h41.6c39.6,0,57,28.3,57,53.6c0,27.5-21.5,53.6-56.8,53.6h-41.8V79.1z M124.3,172.4h24.5c34.9,0,42.9-26.5,42.9-39.7c0-21.5-13.7-39.7-43.7-39.7h-23.7V172.4z}
                 svg{M88.7,56.8c0,5.5-4.5,10.1-10.1,10.1c-5.6,0-10.1-4.6-10.1-10.1c0-5.6,4.5-10.1,10.1-10.1C84.2,46.7,88.7,51.3,88.7,56.8z};
  }
}

\newcommand\orcidicon[1]{\href{https://orcid.org/#1}{\mbox{\scalerel*{
\begin{tikzpicture}[yscale=-1,transform shape]
\pic{orcidlogo};
\end{tikzpicture}
}{|}}}}
\newcommand{\figlabel}[2]{\colorbox{gray!15}{\scriptsize\hspace{#1}#2\hspace{#1}}}

\makeatletter
\newcommand{\pushright}[1]{\ifmeasuring@#1\else\omit\hfill$\displaystyle#1$\fi\ignorespaces}
\newcommand{\pushleft}[1]{\ifmeasuring@#1\else\omit$\displaystyle#1$\hfill\fi\ignorespaces}
\makeatother

\usepackage{cleveref}

\ifCLASSOPTIONcompsoc
  \usepackage[nocompress]{cite}
\else
  \usepackage{cite}
\fi

\begin{document}

\markboth{Journal of XXXX XXXXX XXXX,~Vol.~XX, No.~XX, Month~20XX}%
{N. Surname \MakeLowercase{\textit{et al.}}: Reducing Recurrent Competitive Epidemics and Social Contagions via Dynamic Resource
Allocation}
\title{Reducing Recurrent Competitive Epidemics %
via Dynamic Resource Allocation}

\author{Argyris Kalogeratos, \ 
        Gaspard Abel, \ 
        Stefano Sarao Mannelli%
\IEEEcompsocitemizethanks{%
\IEEEcompsocthanksitem 
A.K. and G.A. are with the Centre Borelli, ENS Paris-Saclay, CNRS, Gif-sur-Yvette, France. G.A. is also with the Centre d'Analyse et de Mathématique Sociales, EHESS, CNRS, Paris, France.%
 S.S.M is with Data Science and AI, Computer Science and Engineering, Chalmers University of Technology and University of Gothenburg, Gothenburg, Sweden, and the School of Computer Science and Applied Mathematics, University of the Witwatersrand, Johannesburg, South Africa.\protect\\
\IEEEcompsocthanksitem 
Correspondence to: A.K.; email: argyris.kalogeratos@ens-paris-saclay.fr.\protect\\
}%
\thanks{Manuscript received December XX, XXXX; revised XXXX XX, XXXX.}}%

\IEEEtitleabstractindextext{%
\begin{abstract}
Motivated by scenarios of epidemic competition, as well as how social %
contagions spread at the level of individuals, this work considers the competition between two conflicting node states that spread over a social graph according to a generic diffusion process.
For this setting, we introduce the \emph{\methodname} (\methodinitials), which is a dynamic resource allocation strategy that favors the preferred state against the other.
Our analysis assumes a generic continuous-time SIS-like (Susceptible-Infectious-Susceptible) diffusion model that allows for: arbitrary node transition rate functions for nodes to change state, and competition between the healthy (positive) and infected (negative) states, which are both diffusive at the same time, yet mutually exclusive at each node. The strategy follows a \emph{minimum-risk-maximum-gain} principle, and its features are particularly relevant for social contagion phenomena. In accordance with the LRIE strategy that we generalize, we show that in this context the \methodinitials strategy remains a greedy solution for the minimization of the number of infected network nodes over time.
Ultimately, simulations are employed to compare the proposed strategy with other existing alternatives, demonstrating that \methodinitials exhibits superior performance across a spectrum of scenarios, including a realistic counter-contagion campaign in a small well-monitored community. %
\end{abstract}

\begin{IEEEkeywords}
Competitive spreading processes, social contagions, behavioral epidemics, epidemic modeling, agent-based models, epidemic control, dynamic resource allocation, %
contact networks.
\end{IEEEkeywords}}

\maketitle

\IEEEraisesectionheading{\section{Introduction}\label{sec:intro}}
\IEEEPARstart{D}{iffusion} processes are being %
studied since the last century, for obtaining models that could help us understand and predict real-world phenomena.
In recent years, the growing amount of available network data led to a surge in the application of diffusion processes, and yielded to a plethora of refined models that are closer to reality.
Diffusion models are used in disparate branches of science: economics (competition among products \cite{winner_takes_all}, viral marketing campaigns \cite{viral-marketing-DP}), epidemiology (disease spreading \cite{marshallComplexSystemsApproach2012}, vaccination \cite{Prakash2010,wu2013dynamical} and allocation of medical resources \cite{valdanoReorganizationNurseScheduling2021,Fekom2019,bakerEpidemicMitigationStatistical2021}), computer and information science (propagation of rumors \cite{twitter2013,rumor2016Zhu,inforumor2018,abelUncoveringSocialNetwork2025}, and 
computer viruses \cite{Hu1318}).

In these last decades, the increasing quantity of data gathered in networked socioeconomic studies has generated a novel field of interest to statisticians, sociologists and physicists. Their approach stems from the idea that various complex social behavior phenomena can be represented and interpreted as \emph{social contagions} \cite{SISa_obesity} that a network of agents can adopt and transmit when via contact with its neighborhood.
Notable examples are: human emotions \cite{SISa_emotions}, behaviors pertaining to public health issues such as obesity diffusion \cite{obesity-DP,SISa_obesity}, smoking cessation \cite{smoking-ces-DP,verma2020optimal}, alcohol abuse \cite{alcohol-ass-DP}, drug use \cite{marshallAdolescentAlcoholUse2014,hoffmanPeerEducatorNetworkHIV2013} or other related to wrongdoing \cite{joseNetworkStructureInfluence2016,calderoni2021recruitment}.

Due to the diffusive nature of such behaviors, adaptive network-level strategies outperform alternatives that are limited to individual-level interventions. This is why researchers, public health professionals, and policy makers should investigate how to leverage social-network-level information %
to optimize the efficacy of counter-contagion interventions.

Many of the above examples correspond to behavioral epidemics with recurrent effects, allowing reinfections, as permanent immunity is never achieved. The Susceptible-Infected-Susceptible (SIS) \cite{NinterSIS2009} model is a good basis to study such epidemics, yet it is deemed simple to describe several real-world phenomena. While several SIS extensions have been developed, we briefly describe a selection of the ones most relevant to the aim of this work.
In terms of model granularity, as opposed to the coarse-grained compartmental models \cite{HethcoteSurvey2000, Newman:2010:NI}, agent-based models (ABMs) provide a framework for high-resolution node interaction.
In this setting a node can become infected according to a transition rate that is \emph{linear} and \emph{local}: \ie the change of a node's state is a stochastic function that depends linearly on the states of its neighbors, hence depicting a social influence effect.
Yet, the assumption for linear infection rates can be restrictive for problems where, for example, the response of individuals exhibits saturation effects \cite{capassoGeneralizationKermackMcKendrickDeterministic1978}. A number of studies have provided evidence supporting the need for more complex transition rate functions to better mimic real-world processes, \eg as in the SISa model \cite{SISa_obesity, SISa_emotions}.
Most such epidemic models consider the infected state to be the only one that diffuses, however, in the context of behavioral epidemics, it is natural to consider that social influence spreads both negative and positive attitudes. As a matter of fact, social norms emerge from the observation of the conduct of influential contacts, %
the adoption and then the emulation of these behaviors, which disseminates them further to others %
\cite{campbellInformalSchoolbasedPeerled2008,joseNetworkStructureInfluence2016,calderoni2021recruitment}.

Across the spectrum of the above application domains, modeling efforts raise   questions about how to reduce an undesirable diffusion.
This line of research has offered numerous intervention techniques (see \cite{epidemicSuppresionSurvey2025} for a survey). Two standard approaches to control an epidemic process is \textit{vaccination} (administration of medicines to specific nodes) and \textit{quarantine} (modification of the network structure by deleting a set of edges).
Most of these strategies are \emph{static} and are based on the network structure, \eg the work in \cite{LRSR} or the greedy algorithm to reduce the spectral radius of a graph from \cite{saha2015approximation}. Additionally, the problem of nonlinear infection rates between nodes is investigated in \cite{kuhlman2013blocking}, where a heuristic is devised for blocking the epidemic spread. There are also \emph{dynamic} strategies that use information about the current state of the network to suggest the best nodes to treat \cite{drakopoulosEfficientCuringPolicy2014a,LRIE,valdanoReorganizationNurseScheduling2021,wang2022competitive}.
Among them, the \emph{Largest Reduction in Infectious Edges} (LRIE) \cite{LRIE} is a greedy algorithm for resource allocation to control SIS epidemics with a limited budget, and is the main source of inspiration for this work.

Alternately, literature in social network analysis has also raised interest in studying social contagion. Namely, opinion dynamic models exhibit similar properties to competitive epidemic models considered here, and optimal strategies to promote a desired state can be as well relevant. Theoretical results on threshold and voting models \cite{lelargeDiffusionCascadingBehavior2012,sadlerDiffusionGames2020,cianfanelliOptimalInterventionsOpinion2025} propose a an optimal intervention strategy to influence the innate opinion of social network agents. Additionally, the epidemic dynamics can also be tackled with a Temporal Point Processes, which rely on probabilistic interaction models. To this end, stochastic optimal control frameworks have been introduced to mitigate the spread of an epidemic \cite{lorch2018stochastic,noorbakhsh2022counterfactual}, or to increase user interaction in social networks \cite{zarezade2018steering}.

The above discussion makes evident that, on the one hand, the literature on modeling one or multiple network diffusion processes has brought a variety of results, including optimal control strategies, while on the other hand, the use of ABMs for the analysis of diverse diffusing social behaviors has gained considerable attention as it could lead to local interventions to facilitate establishing a desired social norm.

\inlinetitle{Contribution}{.}~%
We present the \emph{\methodname} (\methodinitials), a score-based dynamic resource allocation framework for controlling competing recurrent diffusions, seen as Markov processes with arbitrary functions describing node transition rates, \eg exhibiting nonlinearity and saturation effects that are highly relevant to social interactions. %

We focus on social contagions where behaviors can be `healthy' (positive) or `unhealthy' (negative or infected) that are undesired.
The foundation of the \methodinitials strategy is its \emph{minimum-risk-maximum-gain} principle promoting the positive against the negative diffusion in the context of competitive scenarios with conflicting node states (\ie mutually exclusive).
The strategy targets key individuals who pose the greatest transmission risk and, upon recovery, would exhibit a low probability of reinfection while simultaneously amplifying positive diffusion. 
\methodinitials encompasses the LRIE strategy, and as such it constitutes a greedy but more refined way for reducing the number of \emph{infectious edges}, \ie edges connecting an infected and a healthy node and could channel the infection in the network.%

Through rigorous theoretical analysis and extensive simulations on both synthetic and real-world networks, we demonstrate that positive diffusion serves as a potent counter-contagion force, significantly mitigating the complexity of the control problem. 
While this phenomenon is recognized in behavioral studies and health policy \cite{campbellInformalSchoolbasedPeerled2008,hillSpreadingHealthyMood2015,zhangLeveragingSocialInfluence2015}, yet its potential remains largely unexploited within the computational literature. Few studies have addressed the dynamic control of competing diffusion. Specifically, existing social contagion applications are limited to targeting problematic individuals, critically overlooking the capacity of positive diffusion to serve as a effect enhancer against contagion. This work bridges this gap by providing the theoretical and algorithmic framework necessary to operationalize this counter-force.

Finally, we move closer to a realistic counter-contagion campaign through a carefully designed use-case aiming to limit the spread of vaping among adolescents. Our simulated results suggest that intervention strategies can bring valuable gains against social contagion. The discussed control setting requires the administrator to be well-informed about the state of the network nodes, and to be also capable of intervening locally at the node-level. Although those requirements would need to get relaxed for large-scale applications, yet they may not be prohibitive in small well-monitored social environments, such as schools \cite{campbellInformalSchoolbasedPeerled2008,hillSpreadingHealthyMood2015,zhangLeveragingSocialInfluence2015}, penitentiary institutions \cite{dayInterventionsPreventPrison2022,remchEvaluationPrisonViolence2023}, working environments \cite{kensbockEpidemicMentalDisorders2022}, nursing homes \cite{abrahamEffectsGroupInterventions1992,chaoEffectsGroupReminiscence2006}, \etc

\inlinetitle{Organization}{.}~%
In the rest, \Sec{sec:related-work} overviews the literature on competitive SIS models, coupled with a discussion on real-life case studies of behavioral epidemics to which our method is relevant. \Sec{sec:background} focuses on the problem formalization and introduces the set of notations necessary for presenting the \methodinitials algorithm in \Sec{sec:model}. Next, \Sec{sec:results} outlines the experimental protocol and comparative simulations with competitive strategies, including a use-case on reducing vaping in a high school environment. Finally, \Sec{sec:conclusion} concludes this study and gives perspectives for future work.

\begin{figure}[t]\centering\footnotesize
\begin{tikzpicture}[
	node distance=0.6cm and 0.5cm,
	boxq/.style={draw=none, rectangle, align=center, minimum height=0.5cm, minimum width=1cm,fill=red!25},
	box/.style={draw, rectangle, rounded corners, align=center, minimum height=0.9cm, minimum width=1.4cm},
	arrow/.style={->, >=stealth},
	thick,scale=1,
	every node/.style={scale=0.9}
  ]

  \node[box] (start) {SIS \\\cite{NinterSIS2009}};
  \node[boxq, below=of start] (Q1) {\ Competition\ };
  \node[box, below right=1.2 cm of Q1] (Q1no) {SI$_{1+2}$S\\\cite{chen2017fundamental}\,\,\cite{wang2019self}$^*$};
  \node[box, below=0.845cm of Q1] (Q1mix) {SI$_{1|2}$S\\\cite{SI1I2S}};
  \node[box, below left= 1.2 cm of Q1] (Q1yes) {SI$_{1}$I$_{2}$S\\\cite{winner_takes_all}};
  \node[box, below left=1.2cm of Q1yes] (Q1_multi) {SI$_{1}$...I$_{N}$S\\\cite{pare2017multi}$^*$};
  \node[boxq, below right=1.2cm of Q1yes] (Q3) {\ Saturation\ };\
  \node[box, below left= 1.2 cm of Q3] (Q3yes) {\cite{SISa_emotions}\,\,\cite{yangBiVirusCompetingSpreading2018}\\\emph{this work}$^*$};
  \node[box, below right= 1.2 cm of Q3] (Q3no) {\cite{SI1I2S_control1}$^*$\,\cite{sahneh2014competitive}};

   Arrows
  \draw[arrow] (start) -- (Q1);
  \draw[arrow] (Q1) -- node[above right] {no} (Q1no);
  \draw[arrow] (Q1) -- node[above left] {yes} (Q1yes);
  \draw[dashed,->] (Q1) -- node[right] {mixed} (Q1mix);
  \draw[dashed,->] (Q1yes) -- node[above left] {$N$ viruses} (Q1_multi);
  \draw[arrow] (Q1yes) -- node[above] {} (Q3);
  \draw[arrow] (Q3) -- node[above left] {yes} (Q3yes);
  \draw[arrow] (Q3) -- node[above right] {no} (Q3no);

\end{tikzpicture}
\caption{\textbf{Overview of multi-virus SIS models.}~Tree view of the main SIS extensions that consider multiple competing/cooperating infections spreading in a network. * indicates articles that propose epidemic control strategies.}
\label{fig:Tree_SIS}
\end{figure}

\setlength{\fboxsep}{1pt}
\begin{table}[t]
	\centering
	\begin{tabular}{p{1.8cm} p{1.3cm} p{1.5cm} p{1.1cm} p{0.9cm}}
	\cmidrule[0.8pt]{2-5}
	&\multicolumn{4}{c}{\textbf{\methodinitials features}}\\
		\toprule
	\multicolumn{1}{c}{\textbf{Spreading}}	& \multicolumn{1}{c}{\multirow{2}{*}{\textbf{Saturation}}} & \multicolumn{1}{c}{\textbf{Diffusive}}  & \multicolumn{1}{c}{\multirow{2}{*}{\textbf{Control}}} & \multicolumn{1}{c}{\textbf{Network}} \\
	 \multicolumn{1}{c}{\textbf{process}}	&  & \multicolumn{1}{c}{\textbf{healthy state}}  &  & \multicolumn{1}{c}{\textbf{data}} \\
	  \toprule
	\end{tabular}
	\setcellgapes{3pt} \makegapedcells
	\begin{tabular}{p{1.8cm} p{1.3cm} p{1.5cm} p{1.1cm} p{0.9cm}}
	  \raisebox{-1em}{Virus spread}
		& \citewb{InitialViralLoad}\citelgb{yangBiVirusCompetingSpreading2018} & \citelgb{yangBiVirusCompetingSpreading2018}\citelgb{zhangAnalysisStateEstimation2025} & \citewb{InitialViralLoad}\citelgb{yangBiVirusCompetingSpreading2018}$^\star$ \citelgb{zhangAnalysisStateEstimation2025}$^\star$\citelgb{valdanoReorganizationNurseScheduling2021}$^\star$ \citelgb{cristancho-fajardoDynamicResourceAllocation2022}$^\star$ & \citelgb{valdanoReorganizationNurseScheduling2021}\citelgb{cristancho-fajardoDynamicResourceAllocation2022}\\
	  \hline

		Crime, violence\! & \citegb{calderoni2021recruitment}\citegb{joseNetworkStructureInfluence2016}\citewb{remchEvaluationPrisonViolence2023} & \citegb{joseNetworkStructureInfluence2016} & \citegb{calderoni2021recruitment}\citewb{remchEvaluationPrisonViolence2023}& \citegb{joseNetworkStructureInfluence2016}\\
	  \hline

		Positive / Negative sentiment & \citewb{hillSpreadingHealthyMood2015} & \citelgb{SISa_emotions}\citewb{hillSpreadingHealthyMood2015}\citewb{kensbockEpidemicMentalDisorders2022} \citewb{wang2022competitive}\citegb{fanAgentbasedModelEmotion2018} & \citewb{wang2022competitive}$^\star$ & \citewb{hillSpreadingHealthyMood2015}\citegb{fanAgentbasedModelEmotion2018}\\
	  \hline

		\raisebox{-0.5em}{Obesity} & \citelgb{SISa_obesity}\citegb{zhangLeveragingSocialInfluence2015}\citewb{Trogdon2014153} & \citelgb{SISa_obesity}\cite{obesity-DP}\citegb{zhangLeveragingSocialInfluence2015} \citewb{Trogdon2014153} & \citelgb{SISa_obesity}\citegb{zhangLeveragingSocialInfluence2015} \citewb{Trogdon2014153} & \citegb{zhangLeveragingSocialInfluence2015} \\
	  \hline

		\raisebox{-0.5em}{Tobacco use} & \citelgb{verma2020optimal}\citelgb{labzai2018optimal} & \citelgb{verma2020optimal}\citewb{campbellInformalSchoolbasedPeerled2008}\citewb{smoking-ces-DP} \citewb{wymanInfluenceVapingPeer2021}\citewb{valenteSocialNetworkInfluences2023} & \citelgb{verma2020optimal}\citewb{wymanInfluenceVapingPeer2021} \citelgb{labzai2018optimal}$^\star$\citewb{campbellInformalSchoolbasedPeerled2008} & \citewb{campbellInformalSchoolbasedPeerled2008}\citewb{smoking-ces-DP} \citewb{wymanInfluenceVapingPeer2021}\citewb{valenteSocialNetworkInfluences2023}\\
	  \bottomrule
	\end{tabular}
	\smallskip
	\caption{A selection of prior work on real-world epidemic and social contagion use-cases that are relevant to this study. References highlighted with \colorbox{mylightgray}{lightgray} correspond to analytic compartmental models, while \colorbox{mygray}{gray} ones correspond to agent-based models. The rest are data-driven studies. $\star$ indicates articles proposing dynamic control strategies.}%
	\label{tab:tab_applications}%
\end{table}

\section{Motivation and related %
 literature}\label{sec:related-work}
In this section, we motivate both the direction of our technical contributions regarding modeling epidemics as well as the impact of resource allocation in real-world use-cases of social contagion.
\subsection{Modelling competitive epidemics and control interventions}\label{sec:competitive-epidemics}

Coevolving epidemics offer a rich subclass of spreading processes where two or more virus strains try to take on the entire network, possibly one against the others. %
This setting has been mainly studied through extensions of the general SIS setting \cite{NinterSIS2009}, and has then generated a lot of interest in the literature. In comparison, it is only recently that few studies appeared investigating competition for non-recurrent epidemic processes, namely under variations of the Susceptible-Infected-Removed (SIR) model \cite{multi-strain-SIR-2020,competitive-SIR-2022,zhangAnalysisStateEstimation2025}. Regarding the model of our interest, SIS, an indicative (but far from exhaustive) taxonomy of the main multi-virus SIS extensions is illustrated in \Fig{fig:Tree_SIS}, outlining results on the dynamics of spreading processes and their respective outcomes. %
A general epidemic setting is introduced in \cite{SI1I2S}, where two viruses are spreading while interacting, and a parameter $\epsilon$ defines either a competing ($\epsilon<1$) or a cooperating ($\epsilon>1$) nature for the process. A dynamical and linear algebra analysis allowed the determination of the different regimes of existence of one or both viruses in the steady-state. %
For cooperating epidemics, the spread of two co-infecting strains with a phase transition problem is formulated in \cite{chen2017fundamental} and \cite{wang2019self}.
Mutually exclusive epidemics have also been investigated in a variety of settings.
Notably, in the scenario where two viruses spread within a single-layer network, a \emph{winner takes all} phenomenon has been revealed where the virus with higher infectivity not only dominates, but even eradicates the other one \cite{winner_takes_all}. The case of a bi-layer setting, \ie two viruses spreading linearly over different contact networks involving the same set of nodes, is explored in \cite{sahneh2014competitive} and \cite{SI1I2S_control1}; the existence of steady-states of the epidemic where both viruses remain present is proven. This problem is generalized to an arbitrary number of linear spreading processes, SI$_1$...I$_N$S, as well as for time-varying network structures \cite{pare2017multi}.

Extensive research has been carried out on processes with a linear spread, although it has been observed that this assumption often overestimates the dynamics of the epidemic \cite{codella2015agent,hillSpreadingHealthyMood2015}, since they disregard the saturating infectivity. In fact, this type of propagation nonlinearity is the most relevant for social contagion. Regarding this issue, the work in \cite{yangBiVirusCompetingSpreading2018} introduces a competitive SIS model with generic infection rates.

The aforementioned works provide comprehensive analytical insights into the progression of competitive epidemic processes, encompassing both general and advanced perspectives. These results can subsequently be employed in real-life case studies, which is what is discussed in the rest of this section.

\subsection{Applications and use-cases}\label{sec:applications}

ABMs have been employed for modeling health-related behaviors \cite{nianogo2015agent,tracyAgentBasedModelingPublic2018} and network intervention studies \cite{hunter2019social}. \Tab{tab:tab_applications} provides an overview of several diverse application case-studies that have been investigated in the literature. The columns highlight the features that directly relate with this work, namely nonlinear \textit{saturating} epidemic processes where two states are in \textit{competition}, and for which \textit{control interventions} can be applied to a \textit{known network}.

In an epidemiological context, the spread of a disease within a population is approached by a variety of state-based models, each characterizing various compartments such as Susceptible, Exposed, Infected, Quarantined, Recovered, Removed/Dead, \etc The combination of these compartments depends on the properties of the epidemic, such as its time scale, the level of granularity, and other particular circumstances of the outbreak.
Among the existing models, SIS constitutes the foundation for recurrent epidemics.

A direct application concerns interacting diseases, suggesting natural methods for disease control; for example, reducing Dengue transmission by introducing Wolbachia-infected mosquitoes \cite{Mosquito2021,MosquitoBrazil2021}. Computational models have been used to analyze the spread and of these two diseases in \cite{zhang2023optimal}. Additionally, dynamic resource allocation strategies are proposed to tackle problems such as limiting the propagation of diseases in livestock \cite{cristancho-fajardoDynamicResourceAllocation2022}, or examining the infection spread within contact networks between patients and healthcare workers \cite{valdanoReorganizationNurseScheduling2021}.

In a different context, several works have investigated social contagion phenomena related to public health issues, under the scope of competitive spreading processes. First, the sociology of delinquency in adolescence is investigated in \cite{joseNetworkStructureInfluence2016} through school networks, while \cite{calderoni2021recruitment} introduces an ABM to encompass the dynamics of organized crime affiliation, with various intervention techniques. Next, the propagation of positive and negative emotions is tackled by a series of articles:  for the Framingham Heart Study dataset \cite{SISa_emotions}, for the spread of mental health disorders through business networks \cite{kensbockEpidemicMentalDisorders2022}, and for adolescent depression in high schools \cite{hillSpreadingHealthyMood2015}. Obesity has been also a core subject of interest: \eg in the case of adolescents with an ABM \cite{zhangLeveragingSocialInfluence2015}, or in generic networks via statistical analysis \cite{SISa_obesity,obesity-DP,Trogdon2014153}. Similarly, the spread of tobacco use is investigated by quantitative studies \cite{campbellInformalSchoolbasedPeerled2008,smoking-ces-DP,valenteSocialNetworkInfluences2023} and compartmental epidemic models \cite{verma2020optimal,labzai2018optimal}, where the dynamics of the spreading process is studied along with optimal control strategies.

Several real-world pilot studies with deployed intervention experiments have been conducted in high schools. This specific setting is convenient for two reasons: a high school network forms a small `closed' environment, in which the effect of peer influence is particularly important \cite{milburn1995critical,pickeringDiffusionPeerLedSuicide2018}. Indeed, the repeated interaction between individuals is more frequent and intense than in other social contexts. Similarly, environments such as penitentiary institutions \cite{remchEvaluationPrisonViolence2023}, working environments \cite{kensbockEpidemicMentalDisorders2022} and nursing homes \cite{abrahamEffectsGroupInterventions1992} are also settings where social influence plays a key role in shaping behaviors, and this is the reason why they have been also the subject of various awareness-raising campaigns in the last decades in institutions \cite{campbellInformalSchoolbasedPeerled2008,dayInterventionsPreventPrison2022,chaoEffectsGroupReminiscence2006}.

Distinguishing true social influence from environmental factors and homophily remain significant challenges, particularly as network structures evolve alongside behaviors such as delinquency \cite{joseNetworkStructureInfluence2016} or tobacco use \cite{valenteSocialNetworkInfluences2023}. Another important property of the set of applications listed above is that several real-world pilot studies with deployed intervention experiments focus on small `closed' and well-monitored networks, \eg high schools, nursing homes \cite{abrahamEffectsGroupInterventions1992}, and correctional facilities \cite{remchEvaluationPrisonViolence2023}. Unlike open networks, these environments feature high-intensity peer interactions \cite{milburn1995critical,pickeringDiffusionPeerLedSuicide2018}, creating strong amplification effects in diffusion processes. The application of the proposed framework to these specific settings represents a advancement beyond standard computational approaches and constitues a step towards public-health intervention campaigns \cite{campbellInformalSchoolbasedPeerled2008,chaoEffectsGroupReminiscence2006}. Specifically, we provide a quantitative evaluation of how network-aware strategies can exploit this structural intensity to maximize positive behavioral change.

\section{Problem formulation}\label{sec:background}
\subsection{Setup and agent-based diffusion model}\label{sec:setup}
\inlinetitle{Notations}{.} A \emph{graph} $\mathcal{G} \op{=} (\mathcal{V},\mathcal{E})$ is a set of nodes $\mathcal{V}$ representing the individuals in a population, let $N \op{=} |\mathcal{V}|$, which is also endowed with a set of edges $\mathcal{E}\subset \mathcal{V} \times \mathcal{V}$. It can be intuitively represented by its \emph{weighted adjacency matrix} $A \in \real_+^{N\times N}$, where the element $A_{ij} > 0$ if $(i,j) \op{\in} \mathcal{E}$, and $A_{ij} = 0$ otherwise.
We refer to directed graphs without self-loops, \ie $A_{ii}\op{=}0$, $\forall i\op{=}1,...,N$. The \emph{neighborhood} of node $i$ is the set comprising all nodes connected to it with a directed edge, and is denoted by $\mathcal{N}_i \op{=} \{i_k, \forall k \op{\in} \{1,...,d_i\}\!:(i_k,i)\op{\in}\mathcal{E}\}$, where the node degree is $d_i \op{=} \sum_{j} A_{ji}$.
Finally, $\Ind{\cdot}$ is the indicator function.

\inlinetitle{Continuous-time SIS subject to control via node treatments}{.} The
\emph{$N$-intertwined continuous-time homogeneous SIS} model \cite{NinterSIS2009, NinterSIS2011} describes the spread of a disease at each node of a graph. Each node (individual) $i$ can be in either the \emph{susceptible} or the \emph{infected} state: $X_i(t)\op{=}0$ or $1$, respectively. The system at time $t$ is globally characterized by the node state vector $X(t) \op{\in} \{0,1\}^N$.
Adopting the epidemic control aspect from \cite{LRIE, MCM}, with the addition of spontaneous infections, makes the state of node $i$ evolve with the stochastic transition rates:%
\begin{align}\label{eq:SIS}
	X_i(t):\begin{cases}
		0 \rightarrow 1 \quad$ with rate $\quad \alpha + \beta \sum_jA_{ji}X_j(t);\\
		1 \rightarrow 0 \quad$ with rate $\quad \delta + \rho R_i(t),
	\end{cases}
\end{align}
where $\alpha$ is the rate of spontaneous node infections, $\beta$ is the infection aggressiveness, and $\delta$ is the self-healing capacity of a node. The epidemic control is realized by the resource allocation vector $R(t) \op{\in} \{0,1\}^N$, whose coordinate $R_i(t) = 1$ if we treat node $i$ at time $t$, and $0$ otherwise. Finally, $\rho$ is the increase in recovery rate when a node receives a treatment resource. The model is homogeneous as all parameters are the same for all nodes ($\alpha$, $\delta$, $\rho$) or edges ($\beta$); this simplifies the presentation without being a modeling limitation. Moreover, $\alpha$ and $\delta$ can be considered as exogenous factors in the sense that they incur node state changes that are not due to the interactions within the network. %
Modeling bothwise spontaneous changes is particularly relevant in the context of social epidemics, as each individual is subject to both a self-correcting and self-undermining tendencies.

\inlinetitle{A generic two-state recurrent epidemic model}{.} In this work, we introduce the following generic epidemic model.
\begin{definition}\label{def:genericSIS}
Let two {node-specific memoryless functions}: the {infection rate function} $\InfFun_i$ and the $\HealFun_i$ {healing rate function}. A generic two-state recurrent epidemic model
is then defined by the transition rates for each node $i$:
\begin{align}\label{eq:genericSIS}
	\!\!\!X_i(t):\begin{cases}
		0 \rightarrow 1 \quad $ with rate$\quad \InfFun_i(X(t));\\
		1 \rightarrow 0 \quad $ with rate$\quad \HealFun_i(X(t))  + \rho R_i(t),
	\end{cases}
\end{align}
which induces a Markovian Poisson process with rate:
	\begin{equation*}
		\lambda_i(t) = \Ind{X_i(t)=0}\,\InfFun_i(X(t)) + \Ind{X_i(t)=1}\,\HealFun_i(X(t)).
  \end{equation*}
\end{definition}
Since the administrator wishes to reduce the infection which has negative effects, we also call $\InfFun_i$ and $\HealFun_i$ as \emph{positive/negative diffusion functions}. Those depend on the current network state $X(t)$ and implicitly on the network structure (this is omitted from our notation). In our context, it is natural to additionally assume that $\InfFun_i$ and $\HealFun_i$ are monotonous \wrt $X(t)$: $\InfFun_i$ increases with $X(t)$ and $\HealFun_i$ %
increases with $\notX(t)$.

This SIS variant %
inherits the control mechanism from \Model{eq:SIS}, namely the allocation of resources having stochastic treatment effect on individual nodes. Although this is not in the scope of this work, we may note that \Model{eq:genericSIS} suggests that control could also be attempted through global or local tuning of the $\InfFun_i$ or/and $\HealFun_i$ function shapes.
\subsection{Dynamic resource allocation
} \label{sec:greedyDRA}
Dynamic interventions are meaningful when the the time-scale needed for the interventions to take effect is comparable to the speed of the epidemic (\ie node state transitions).
In the \emph{Dynamic Resource Allocation} (DRA) problem \cite{LRIE, MCM}, an SIS model variant (\Model{eq:SIS}) is admitted, and the objective is to administer a budget of $\budget$ treatment resources (units), each having strength $\rho$, in order to suppress an undesired state that diffuses. The treatments cannot be stored for later use ($\sum_i R_i(t) \leq b$, $\forall t$), and their effect is limited as a node can receive at most one treatment ($R_i(t) \in \{0,1\}$, $\forall t,i$). Dynamic score-based strategies, which invoke the resource re-allocation procedure in \Alg{alg:score-based} whenever intervention is possible, can provide efficient solutions to the DRA problem and are rather simple to implement. A greedy dynamic score-based strategy is developed in \cite{LRIE}, called \emph{Largest Reduction in Infectious Edges} (LRIE). Each node is assigned a score that quantifies its criticality for the future spread of the infection. %

Other score-based solutions have been proposed, \eg based on fixed Priority Planning \cite{MCM}, or static ones based on spectral analysis \cite{LRSR,saha2015approximation} (see details in \Sec{sec:results}). In a complementary research direction, such scoring functions have also been used in Restricted- and Sequential-DRA settings \cite{Fekom2019}, which constrain the administrator so she can access and intervene only to few nodes through resource allocation.

\begin{algorithm}[t]\small
	\caption{Generic score-based DRA control strategy$^\star$}%
	\begin{algorithmic}[1]
	\Require: adjacency matrix $A$; network state $X(t)$ at time $t$; budget available $\budget$
	\Ensure: resource allocation vector $R(t)$, with $1$ for targeted nodes and $0$ otherwise, while respecting the budget $\sum_{i=1}^N R_i(t) \leq \budget$
	\vspace{0.4mm}
	\hrule
	\vspace{0.7mm}
	\For{$i = 1,...,N$}
		\State $S_i \leftarrow \texttt{computeScore} (i,X(t),A)$
	\EndFor
	\State $nodeRanking \leftarrow \texttt{sort}(S)$
	\State $R(t) \leftarrow \texttt{selectTop}(nodeRanking,\!\ \min(\budget,\sum_iX_i(t)))$
	\State \Return $R(t)$
	\end{algorithmic}
		\hrule %
		\vspace{1mm}
\noindent{\normalsize $^\star$}\,{\footnotesize Resource re-allocation occurs each time the control strategy gets invoked for intervening to the spreading process, and the resources remain in effect on the target nodes until a next update.}
	\label{alg:score-based}
\end{algorithm}
\section{\methodname: analysis and algorithm}\label{sec:model}
We propose the DRA strategy named \emph{\methodname} (\methodinitials) that at each time identifies and treats in a greedy fashion the most critical nodes in order to reduce the epidemic as quickly as possible. The gravity of the epidemic can be expressed in terms of the number of infected nodes at time $t$, denoted by $N_I(t) = \sum_i X_i(t)$. The approach generalizes the LRIE strategy \cite{LRIE} to a broader class of epidemics, featuring also competition, which is the main interest of this work. In a Markovian setting, given the fixed characteristics denoted by $\Lambda$ (\ie the graph structure and the characteristics of the propagation), and the instantaneous state of the population $X(t)$, the best intervention expressed by the resource allocation vector $R$ in \Model{eq:SIS} and \Model{eq:genericSIS} that would minimize the cost function:
\begin{equation}\label{eq:minfuncA}
	\!\!\!\!C(\Lambda, X, \gamma) = \int_{u=0}^{\infty} \!e^{-\gamma u} \Exp[N_I(t\op{+}u)\padded{|}X(t)\op{=}X]\,du.\!\!
\end{equation}
$\Lambda$ remains fixed throughout the process, so we omit it from the rest of the notations. The cost function is parametrized by $\gamma  \in \real_+$ that can be chosen so as to give emphasis on short-term effects. Considering $\Phi_{t,X}(u) = \, \Exp[N_I(t\op{+}u)\padded{|}X(t)\op{=}X]$ and then taking the Taylor series expansion \wrt $u$ at $0$, yields:
\begin{equation}\label{eq:minfuncB}
		\!\!\!\!\!\!(\text{\ref{eq:minfuncA}}) = \frac{1}{\gamma}\Phi_{t,X}(0) + \frac{1}{\gamma^2}\Phi_{t,X}^{'}(0) + \frac{1}{\gamma^3}\Phi_{t,X}^{''}(0) + \bigO\Big(\frac{1}{\gamma^4}\Big).\!\!\!
\end{equation}

Next, we present the final results of the derivation of each of the three first terms, while their detailed evaluation can be found in \Appendix{sec:appendix}. As treatments %
are always administered to infected individuals, let us concentrate at an infected node $i$. %
At this point, it is convenient to define the generic rate function $\FFun \in \{\HealFun, \InfFun\}$ to refer to both the positive or the negative diffusion, and special notations for the updates on $\FFun_j$, $j \neq i$, in the hypothetical case where node $i$ recovers, and that recovery to be the only change occurred in the network:
\begin{align}\label{eq:rate-updates}
	\begin{aligned}
	\FFun_j^{-i} & = \FFun_j(X_1,...,X_i\op{=}0,...,X_N),\\
	\DeltaF{j}{i} & = \FFun_{j} \op{-} \FFun^{-i}_j.%
	\end{aligned}
\end{align}
In this sense, \Eq{eq:rate-updates} produces the four definitions: $\forall j \neq i$, $\HealFun_j^{-i}$, $\InfFun_j^{-i}$, $\DeltaH{j}{i} = \HealFun_j - \HealFun_j^{-i} \op{\leq} 0$, and $\DeltaI{j}{i} = \InfFun_j - \InfFun_j^{-i}\op{\geq} 0$. The last two give the difference in the rates after a hypothetical recovery of node $i$. %
Their signs are so due to $\HealFun_i$ (resp. $\InfFun_i$) being monotonically decreasing (resp. increasing) with the infection level $X(t)$, and the fact that any node recovery essentially reduces $X(t)$. %
Therefore, the final forms of the derivatives appearing in \Eq{eq:minfuncB} become:
\begin{align}
	\Phi_{t,X}(0) & = \sum_i X_i, \\
	\Phi_{t,X}^{'}(0) & = -\sum_i \HealFun_i X_i -\rho \sum_i R_i X_i + \sum_i \InfFun_i \,\notX[i],\\
	\begin{split}
	\Phi_{t,X}^{''}(0)& =\Xi (t) + \rho \sum_i  X_i R_i\bigg\{(\HealFun_i+\InfFun_i)+
	\\
	&\quad\quad\quad+\sum_{j\not=i} \Big[X_j(\DeltaH{j}{i})-\notX[j](\DeltaI{j}{i})\Big] \bigg\},
	\end{split}
	\label{eq:3rd term}
\end{align}
where we let $\notX_i = 1-X_i$, and $\Xi(t)$ to be a function  that absorbs any terms that are independent to $R_i$. The terms of the expansion describe the effects of healing a node (\ie after it recovers) on the network.
The first two %
trivially suggest treating only infected nodes.
The third one (second order term), though, quantifies the contribution of managing to heal a specific node to the minimization of the cost function.
According to \Eq{eq:3rd term}, we derive the following \methodinitials score for each infected node $i$:
\begin{equation}\label{eq:genericScore}
S_i = -\bigg[ \xoverbrace{\tikz[baseline]{\node(d4) {$(\HealFun_i$}} + \tikz[baseline]{\node(d5){$\InfFun_i)$}}\!\!}^\text{\phantom{p}effect on $i$\phantom{p}} + \overbrace{\sum_{j\not=i}\Big[\tikz[baseline]{\node(d6) {$X_j(\DeltaH{j}{i})$}}-\tikz[baseline]{\node(d7){$\notX[j](\DeltaI{j}{i})$}}\,\Big]\phantom{\bigg(\bigg)}\!\!\!\!\!\!\!\!\!\!\!\!\!}^\text{effect of the supposed healed $i$ on the network%
}\!\bigg]\!.
\end{equation}
\begin{tikzpicture}[remember picture,overlay]
    \draw[OliveGreen,thick,->] (d4) to [in=90,out=245] + (240:.9cm) node[anchor=north,text = black,text width=3cm,align=center] {\scriptsize{self-healing and\\\vspace{-0.5em}  social healing}};
    \draw[orange,thick,->] (d5) to [in=90,out=265] +(295:.86cm) node[anchor=north,text = black] {\scriptsize{vulnerability}};
    \draw[red!30!blue!70,thick,->] (d6) to [in=90,out=265] +(285:.8cm) node[anchor=north,text = black,text width=4cm,align=center]
     {\scriptsize{contribution to\\\vspace{-0.5em} social healing}};
    \draw[red,thick,->] (d7) to [in=90,out=265] +(275:.8cm) node[anchor=north,text = black,text width=3cm,align=center]
     {\scriptsize{virality}};
\end{tikzpicture}

\vspace{0.7cm}
\inlinetitle{Interpretation}{.}~To discuss the rationale of the score in \Eq{eq:genericScore}, we distinguish two parts. The first is the quantification of node $i$'s total transition rate, $\HealFun_i \op{+} \InfFun_i$: this says that if $i$ could get easily reinfected (high $\InfFun_i$) or/and could anyway recover rapidly due to self-healing and positive diffusion that we call ``social healing'' (high $\HealFun_i$), then it is not a good candidate to invest resources on. The second part quantifies the effect that a supposed recovered $i$ would bring to the network: by recalling that always $\DeltaH{j}{i} \op{\leq} 0$ and $\DeltaI{j}{i} \op{\geq} 0$, we can see that \methodinitials assigns higher scores to prioritize nodes that would bring a high gain in social healing (high negative $\DeltaH{j}{i}$ value for infected $j$) and a big drop of infectivity (high positive $\DeltaI{j}{i}$ value for healthy $j$).
\Fig{fig:simple-example} illustrates how the various factors related to the positive and the negative diffusion are taken into account when computing the \methodinitials scores.

Notice that the locality of the \methodinitials score is determined by the definition of the functions $\HealFun_i$ and $\InfFun_i$, and can potentially depend on the whole network.
In the next section, we introduce assumptions to arrive to a variant of this score that is simpler and more useful in practice.

\begin{figure*}[t]\small\centering
	\newcommand{\labelA}{{\text{A}}}
	\newcommand{\labelB}{{\text{B}}}
	\newcommand{\labelC}{{\text{C}}}
	\newcommand{\labelD}{{\text{D}}}
	\newcommand{\labelE}{{\text{E}}}
	\newcommand{\redColor}{red!53}
	\newcommand{\greenColor}{red!40!green!60}
	\newcommand{\orangeColor}{orange}
	\newcommand{\blueColor}{red!30!blue!70}
	\newcommand{\elw}{1.2}
	\newcommand{\nsz}{0.75}
	\newcommand{\DennisX}{0.5} \newcommand{\DennisY}{2.5}
	\newcommand{\AliceX}{2.75}     \newcommand{\AliceY}{2.5}
	\newcommand{\ClaireX}{4.5}\newcommand{\ClaireY}{4.5}
	\newcommand{\BobX}{4.5}  \newcommand{\BobY}{0.5}
	\newcommand{\ErikaX}{2.5}   \newcommand{\ErikaY}{7}
		\centering
\hspace{2em}%
	\begin{minipage}{0.3\textwidth}
	\subfigure[Contact network (straight black edges) and factors affecting node scores (colored arcs).]{%
		\scalebox{1.05}{%
		\begin{tikzpicture}%
			arrow/.style={->, >=stealth},
		\clip (0,0) rectangle (5,5.5);
			\Vertex[size=\nsz,x=\DennisX,y=\DennisY,color=\greenColor,style={very thick,draw=\redColor},label=Dennis]{d}
		\Vertex[size=\nsz,x=\BobX,y=\BobY,color=\redColor,,style={very thick,draw=\greenColor},label=Bob]{b}
		\Vertex[size=\nsz,x=\ClaireX,y=\ClaireY,color=\redColor, style={very thick,draw=\greenColor},label=Claire]{c}
		\Vertex[size=\nsz,x=\AliceX,y=\AliceY,color=\redColor,,style={very thick,draw=\greenColor},label=Alice]{a}
    \Edge[color=\blueColor,bend=-45,Direct=true,lw=\elw,label=$\DeltaH{\labelB}{\labelA}$](b)(a)
    \Edge[color=\blueColor,bend=-50,Direct=true,lw=\elw,label=$\DeltaH{\labelA}{\labelB}$](a)(b)
		\Edge[color=\redColor,bend=-15,Direct=true,lw=\elw,label=$\InfFun_\labelA$](a)(b)
    \Edge[color=\redColor,bend=-15,Direct=true,lw=\elw,label=$\InfFun_\labelA$](a)(c)
    \Edge[color=\blueColor,bend=-55,Direct=true,lw=\elw,label=$\DeltaH{\labelA}{\labelC}$](a)(c)
    \Edge[color=\blueColor,bend=-45,Direct=true,lw=\elw,label=$\DeltaH{\labelC}{\labelA}$](c)(a)
    \Edge[color=\redColor,bend=-15,Direct=true,lw=\elw,label=$\InfFun_\labelC$](c)(a)
		\Edge[color=\redColor,bend=-15,Direct=true,lw=\elw,label=$\InfFun_\labelB$](b)(a)
    \Edge[color=\orangeColor,bend=-45,Direct=true,lw=\elw,label=$\DeltaI{\labelD}{\labelA}$](a)(d)
		\Edge[color=\greenColor,bend=-15,Direct=true,lw=\elw,label=$\HealFun_\labelD$](d)(a)
    \Edge[color=\redColor,bend=-15,Direct=true,lw=\elw,label=$\InfFun_\labelA$](a)(d)
		\Edge[,style={dashed},](a)(d)
    \Edge[,,](b)(a)
    \Edge[,,](c)(a)
		\end{tikzpicture}%
			}%
			\label{fig:scoring-variables}%
	}%
	\end{minipage}
	\hspace{3em}%
	\begin{minipage}{0.60\textwidth}%
	\vspace{10.2em}
	\subfigure[The factor values used to compute the \methodinitials and LRIE scores.]{\footnotesize%
		\begin{tabular}{p{0.9cm} p{0.5cm} p{0.5cm} p{1.2cm} p{1cm} p{1.5cm} p{1cm}}%
		  \toprule
		  \textbf{Node} $i$ & $\HealFun_i$ & $\InfFun_i$ & $\sum_j\DeltaH{j}{i}$ & $\sum_j\DeltaI{j}{i}$ & \ \ \ $S_i^{\text{\methodinitials}}$& $ \ S_i^{\text{LRIE}}$ \\
		  \toprule
		  Alice & $\delta$ & $2\beta$ & $-2\gamma$ & $\beta$ & \textcolor{blue}{$-\delta - \beta + 2\gamma$} & \textcolor{orange}{$-\delta - \beta$} \\
		  Bob & $\delta$ & $\beta$ & $-\gamma$ & 0 & $ -\delta- \beta + \gamma$ & \textcolor{orange}{$-\delta - \beta$} \\
		  Claire & $\delta$ & $\beta$ & $-\gamma$ & 0 & $ -\delta- \beta + \gamma$ & \textcolor{orange}{$-\delta - \beta$} \\
			Dennis & -- & -- & -- & -- & -- & -- \\
		  \bottomrule
			\vspace{0.5em}
		\end{tabular}
			\label{tab:tab_scoring_example}
	}
	\end{minipage}
	\caption{\textbf{Minimal example comparing \methodinitials and LRIE scores.}~a)~A simple undirected and unweighted contact network between $4$ individuals, $3$ of which are infected. The edges between agents are shown as straight black lines, out of which those being dashed are the \emph{infectious edges} that connect nodes in different states and hence can channel both positive and negative diffusion.
	The factors involved in %
	\methodinitials score (\Eq{eq:genericScore}) appear as labeled arcs. Node boundaries, green for infected and red for healthy nodes, indicate the spontaneous self-healing and self-infection, respectively. b) Table with the values of the factors involved in the \methodinitials and LRIE scores for an infected node $i$, in the case of linear infection. The parameters $\delta$, $\gamma$, and $\beta$ are the self-healing, social healing, and social infection rates, respectively. The \methodinitials score prioritizes Alice over Bob and Claire, due to the added contribution of social healing she could bring when she would have recovered, while LRIE treats them equally. Dennis being healthy, he is not considered for receiving a treatment.}\label{fig:simple-example}
\end{figure*}

\subsection{Simplified \methodinitials score}\label{sec:appendix2}
The score-based strategy suggested by \Eq{eq:genericScore} %
can be simplified under three rather standard assumptions in SIS modeling, concisely expressed in terms of the generic rate function $\FFun \in \{\HealFun, \InfFun\}$ (see \Eq{eq:rate-updates}).
\begin{itemize}[leftmargin=0.5cm]
\item[$\mySqBullet$] \textbf{Assumption~1~--~\emph{Locality}:}~%
	The transition rate of each node $i\in \mathcal{V}$ depends only on its neighbors $i_1,..., i_{d_i} \in \mathcal{N}_i$%
:
	\begin{align}
		\begin{aligned}
			\FFun_i = \FFun_i(X_{i_1},...,X_{i_{d_i}}).
		\end{aligned}
	\end{align}
\item[$\mySqBullet$] \textbf{Assumption~2~--~\emph{Exchangeability}:}~%
	The transition rate of each node $i\in \mathcal{V}$ is invariant to the node permutations ($\pi$) %
	of its neighbors: %
	\begin{align}
		\begin{aligned}
		\!\!\!\!\!\!\!\!\!\!\!\!\FFun_i(X_{i_1},...,X_{i_{d_i}})=\ \FFun_i(X_{\pi(i_1)},...,X_{\pi(i_{d_i})}).\\
		\end{aligned}
	\end{align}
\item[$\mySqBullet$] \textbf{Assumption~3~--~\emph{Homogeneity}:}~%
	The transition rate functions are the same for all nodes $i\in \mathcal{V}$:
	\begin{align}
		\begin{aligned}
		\FFun_i &= \FFun, \ \ \forall i\in\{1,...,N\}.%
		\\
		\end{aligned}
	\end{align}
\end{itemize}

Assumptions\,1-2 restrict the class of rate functions to those that only depend on a node's neighborhood state and size. Indeed, given the node degree $d_i$, the state of its neighborhood $\{X_{i_1},...,X_{i_{d_i}}\} \op{\in} \{0,1\}^{d_i}$, and the number of infected neighbors $n_i = \sum_jA_{ji}X_j$, we have for all $i\in \mathcal{V}$:
\begin{align}
    \!\!\!\FFun_i(X_1,...,X_{d_i}) = \FFun_i(\overbrace{1,...,1}^{n_i},\overbrace{0,...,0}^{d_i-n_i}) =: \FFun_i(n_i,d_i).%
\end{align}

Due to Assumption\,1, for any node $j$ non-adjacent to $i$, it holds
$%
\forall j\op{\not\in}\mathcal{N}_i : \DeltaF{j}{i} \op{=} \DeltaF{i}{j} \op{=} 0$.	Moreover, as Assumption\,2 renders neighbors indistinguishable, we can eliminate the reference to node $i$ from the notations in \Eq{eq:rate-updates}:
$\forall i \op{\in} \mathcal{N}_j : \FFun_j^- \op{\equiv} \FFun_j^{-i}$ and $\DeltaF{j}{} \op{\equiv} \DeltaF{j}{i}$. %
Let us bringing the explicit $\HealFun_i$ and $\InfFun_i$ back in the discussion. According to Assumption~3, we can consider that the same positive and negative transition rate functions act on all the network nodes, which writes:
\begin{align}\label{eq:abstractFunctions}
	\begin{aligned}
		\InfFun & = \ \InfFun(n_i,d_i),\\
		\HealFun & = \HealFun(n_i,d_i).
	\end{aligned}
\end{align}
The above assumptions and discussion lead to Remarks \ref{remark: SIS}-\ref{remark:LRIE-gLRIE-equiv}.

\begin{remark}\label{remark: SIS}
\textbf{Encompassing the standard SIS model.}~
Given Assumptions\,1-3, the standard SIS \Model{eq:SIS} can be recovered by the introduced \Model{eq:abstractFunctions} by setting:
\begin{align}\label{eq:recover-SIS}
	\begin{aligned}
		\InfFun(n_i,d_i) &= \beta{\textstyle \sum_j}A_{ji}X_j \op{=} \beta n_i,\\
		\HealFun(n_i,d_i) &= \delta.
	\end{aligned}
\end{align}
\end{remark}
\begin{remark}\label{remark:explicitScore}
\textbf{Simplified \methodinitials score.}~Assumptions\,1-3 yield the following simplified variant of the \methodinitials score:
\begin{equation}\label{eq:explicitScore}
	\!\!\!S_i = -\bigg[ (\HealFun_i+\InfFun_i)+\sum_j A_{ij} \Big[X_j(\DeltaH{j}{})-\notX[j](\DeltaI{j}{})\Big]\bigg].\!\!
\end{equation}
Compared to the more generic \Eq{eq:genericScore}, this score is local as the transition functions and the summation look only at the neighborhood of node $i$ (recall also that the adjacency matrix $A$ has a zero diagonal). %
\end{remark}

\begin{remark}\label{remark:LRIE-gLRIE-equiv}
\textbf{Equivalence of \methodinitials and LRIE score rankings.}~Under the standard SIS \Model{eq:SIS} without competition, the scoring function gets a form as in \Remark{remark: SIS}, with: %
\begin{align}\label{eq:gLRIE-LRIE-paameters}
	\begin{aligned}
		\InfFun_i &= \alpha + \beta{\textstyle \sum_j}A_{ji}X_j,\\
		\HealFun(n_i,d_i) &= \delta, \\
		\DeltaH{j}{} &= 0, \\
		\DeltaI{j}{} &= \beta.
		\end{aligned}
\end{align}
Then, the simplified \methodinitials score (\Eq{eq:explicitScore}) reduces to:
	\begin{equation}\label{eq: score LRIE}
		S_i = -\bigg[(\alpha + \delta) + \beta\sum_jA_{ji}X_j-\beta\sum_j A_{ij} \,\notX[j]\bigg],
\end{equation}
which is equivalent to the LRIE score up to the constant term $\alpha + \delta$ that does not affect node ranking.
\end{remark}
\begin{figure*}\small
	\centering
	\figlabel{5pt}{\scriptsize\myfont \emph{{Scenario without competition}} ($\HealFun = 0$)}
	\makebox[\linewidth][c]{
		\includegraphics[width=\linewidth]{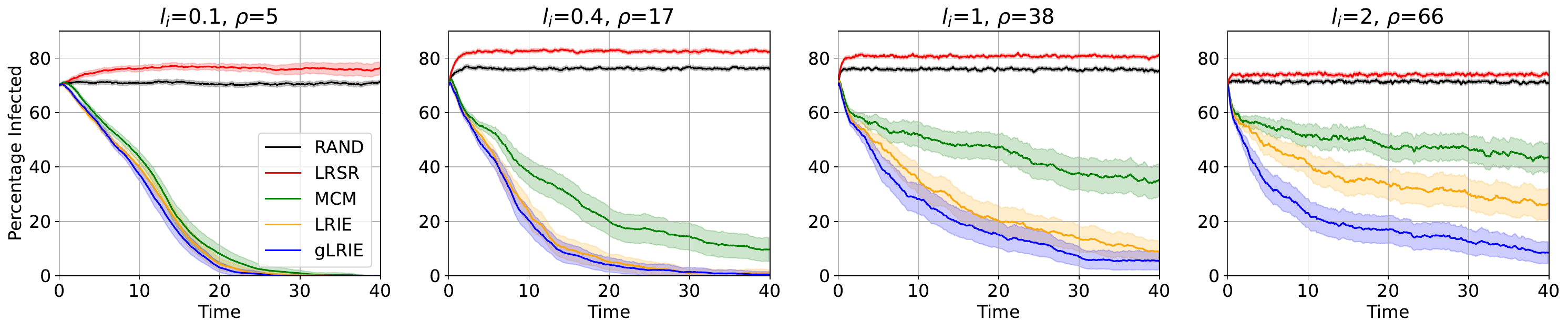}
	}
	\beforecaptvskip
	\caption{\textbf{Evolution of a single epidemic from linear to nonlinear spreading for different intervention strategies.}~The percentage of infected nodes over time in \Erdos\ graphs of $300$ nodes with average degree $8$, when different strategies are employed. %
    \emph{Only the negative diffusion is considered} ($\HealFun\op{=}0$). Each plot shows an average over $1000$ simulations of the generalized SIS \Model{eq:genericSIS} using \Eq{eq:spreading_absolute}. From (a) to (d), the diffusion moves from being essentially linear to nonlinear. The saturation level is fixed to $s_{_\InfFun}\op{=}10$, and the number of resources to allocate is $\budget\op{=}10$.}
	\label{fig:seq}
\end{figure*}

\inlinetitle{Algorithm and computational complexity}{.}
At time $t$, the \methodinitials strategy takes as input the network state $X(t)$ and the budget of resources $\budget$; it computes independently the criticality score for each node (\Eq{eq:genericScore}); ranks the nodes according to their scores; finally, it treats as many nodes as the budget allows, \ie $\sum_i R(t) \op{=} \min(\budget,\sum_iX_i(t)))$.
The simplified \methodinitials scoring function (\Eq{eq:explicitScore}) reduces the computational complexity of the strategy (\Alg{alg:score-based}) to $\bigO(Nd_{\text{max}} \op{+} N\log N)$, where %
$d_{\text{max}}$ is the maximum node degree. %

\inlinetitle{Discussion}{.}~%
The main contribution of this paper is the generic \methodinitials scoring strategy that relies on a SIS model reformulation, and is suitable for %
multiple recurrent spreading processes in a network. \methodinitials bears a number of properties, some of which are inherited by the DRA formulation and its greedy approach, and therefore are shared with the included LRIE, while others stem from the introduction of the generic positive and negative diffusion functions:
\begin{itemize}[leftmargin=0.5cm]
		\item \emph{Generic transition rates}: To capture the complexity of social interactions, %
		both the positive and negative diffusion rates are designed to be generic, exhibiting saturating effects and exogenous intensity to assimilate social influence outside the network, and can potentially vary for each individual.
    \item \emph{Distributivity}: As a strategy relying on local scores, \methodinitials allows independent score computations at individual nodes, and requires only a centralized score ranking. %
    \item \emph{Adaptivity}: It exhibits adaptability to changes of the network structure, which is relevant in scenarios involving dynamic networks emerging in behavioral epidemics, or in controversial public debates that frequently trigger dramatic changes in the contact network in short time.
    \item \emph{Edge directionality}: %
		Taking into account edge directionality is essential in the context of social contagion. This is taken into account to ensure comprehensive coverage.
\end{itemize}

\section{Simulations}\label{sec:results}

In this section, we first present our experimental setup, the specific diffusion functions we choose for \methodinitials, the competitors we choose from the literature. Next, we present simulations on random and real networks that validate the core properties of \methodinitials. Finally, we design a semi-synthetic experiment that illustrates the applicability of our approach to real-world campaigns.
\subsection{Experimental setup}\label{sec:exp_setup}
\inlinetitle{Diffusion functions}{.}
We employ the generalized SIS \Model{eq:genericSIS}. In order to encompass particular aspects in social behaviors, such as \emph{nonlinearity} and \emph{saturation} in the node transition rates, we consider sigmoid functions:
\begin{align}\label{eq:spreading_absolute}
\begin{aligned}
		\ \InfFun(n,d) &= \alpha + s_{_\InfFun} \bigg[1-\frac{2}{1+\exp\left({2\slopeSymbol_{_\InfFun} \frac{n}{\xi}}\right)}\bigg];\\
		\HealFun(n,d) &= \delta + s_{_\HealFun} \bigg[1-\frac{2}{1+\exp\left({2\slopeSymbol_{_\HealFun} \frac{(d-n)}{\xi}}\right)}\bigg],
\end{aligned}
\end{align}
where $\saturationSymbol_{_\InfFun}$ (resp. $\saturationSymbol_{_\HealFun}$) parameter controls the saturation level and $\slopeSymbol_{_\InfFun}$ (resp. $\slopeSymbol_{_\HealFun}$) the slope at the origin.
Recall that $n$ is the number of infected neighbors, $d$ is the node degree, while $\xi \in \{1, d\}$ is a normalization term that makes the state transition rate of a node dependent to either the number ($\xi=1$) or the fraction ($\xi=d$) of its neighbors being in the opposing state.
Without loss of generality, we henceforth set $\xi=1$ that %
takes explicitly into consideration the fact that individuals' limited total capacity may allow social interaction with a small number of their contacts each time.
The above equations are expressed for the case of an unweighted and undirected graph. For a directed graph without weights, $d$ would be replaced by the node in-degree $d_{\text{in}}$. For a weighted graph, $n$ could be replaced by the weighted sum of infected neighbors ($\sum_j A_{ji} X_j$), and $d$ by the weighted sum of all neighbors ($\sum_j A_{ji}$).
The functions $\InfFun$ and $\HealFun$ are repetitively evaluated on a finite and discrete set of points, $(n,d) \in \{0,1,...,d_\text{max}\}^2$, %
therefore, they could be precomputed in two matrices %
of size $(d_\text{max}+1) \times (d_\text{max}+1)$ each, and retrieved in constant time during simulations.

\inlinetitle{Other strategies}{.}
As a naive baseline, we use the Random Allocation (RAND) that targets infected nodes at random. The second competitor is the Largest Reduction in Spectral Radius (LRSR) \cite{LRSR}, which is based on spectral graph analysis generalized to arbitrary healing effects ($\rho \op{\not=} \infty$). LRSR selects nodes that maximize the eigen-drop of the largest eigenvalue of the adjacency matrix, known as \emph{spectral radius}. Next comes the MaxCut Minimization (MCM) \cite{MCM}, which uses the \emph{priority planning} approach. The strategy proceeds according to a precomputed node \emph{priority-order} that is a linear arrangement of the network with minimal \emph{maxcut}, \ie maximum number of edges need to be cut in order to split the ordering in two parts. \footnote{The comparison with the CURE algorithm \cite{drakopoulosEfficientCuringPolicy2014a} is omitted, as in \cite{MCM} it has been shown to have inferior and less stable performance compared to MCM.} The last most direct competitor is the greedy dynamic LRIE strategy \cite{LRIE} that is generalized in this work.

\begin{figure}[t] \footnotesize
	\centering
	\figlabel{3pt}{\scriptsize\myfont \emph{{Scenario without competition}} ($\HealFun = 0$)}\ %
		\figlabel{3pt}{\scriptsize\myfont \emph{{Scenario with competition}} ($\HealFun \neq 0$)}

	\vspace{1em}
	\scriptsize\myfont \emph{\Erdos~(ER)}\\
	\vspace{-1mm}
	\makebox[\linewidth][c]{%
	\subfigure{
		\includegraphics[clip=true, width=\linewidth]{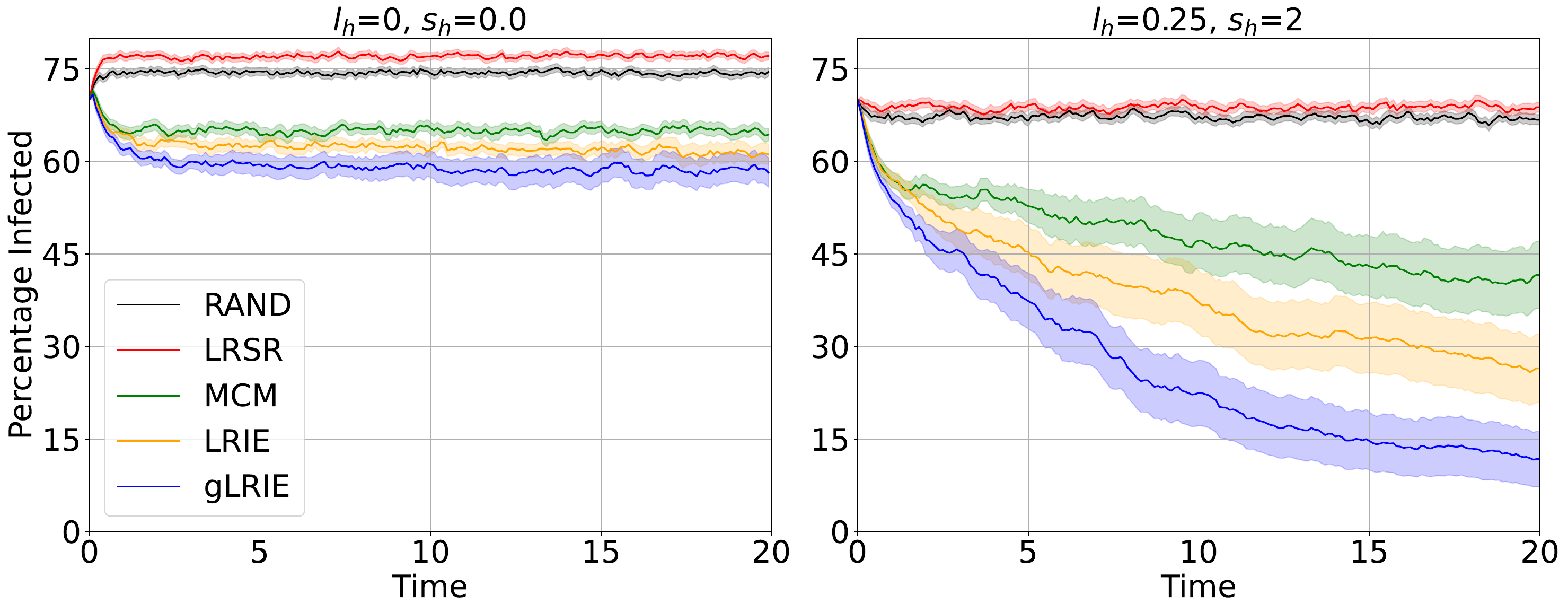}
	}}
	\\
	\centering
	\vspace{2mm}
	\scriptsize\myfont \emph{Preferential Attachment (PA)}\\
	\vspace{-1mm}
	\makebox[\linewidth][c]{%
	\subfigure{
		\includegraphics[clip=true, width=\linewidth]{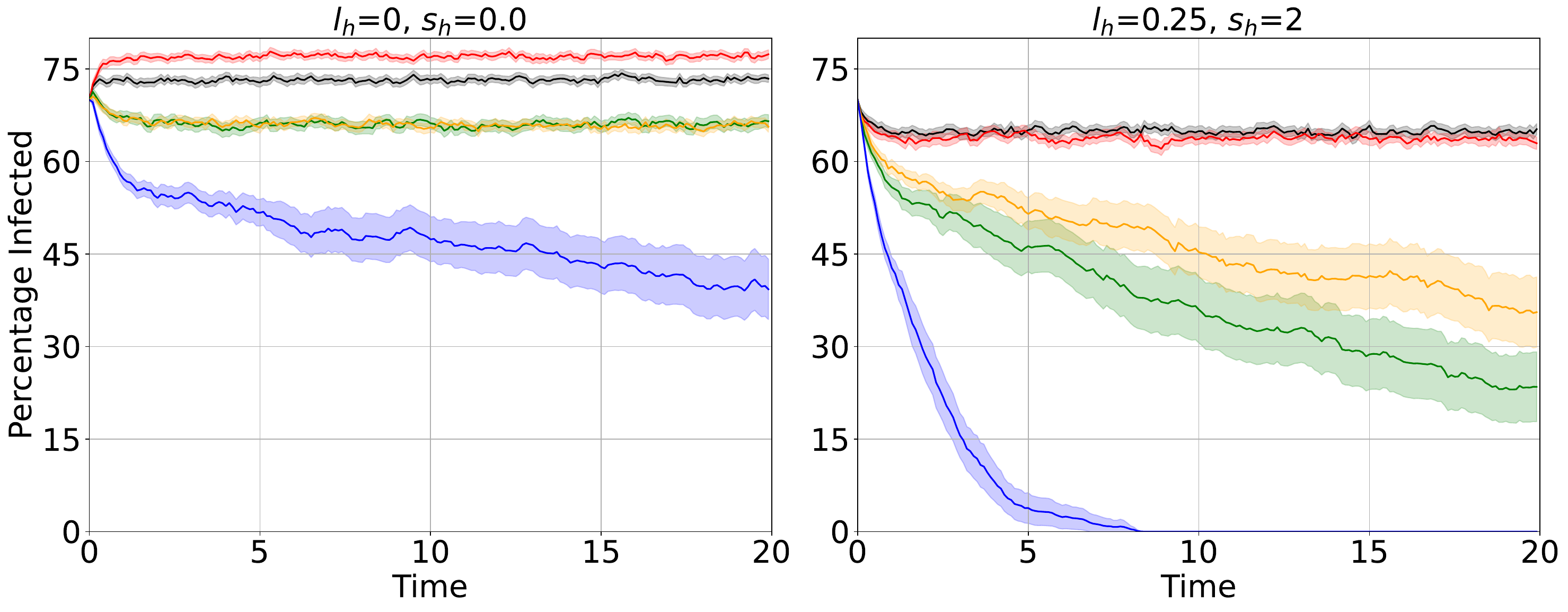}
		\label{fig:prefAtt_noHD}
	}}
	\\
	\centering
	\vspace{2mm}
	\scriptsize\myfont \emph{Hierarchical \Erdos\ (hierER)}\\
	\vspace{-1mm}
	\makebox[\linewidth][c]{%
	\subfigure{
		\includegraphics[clip=true, width=\linewidth]{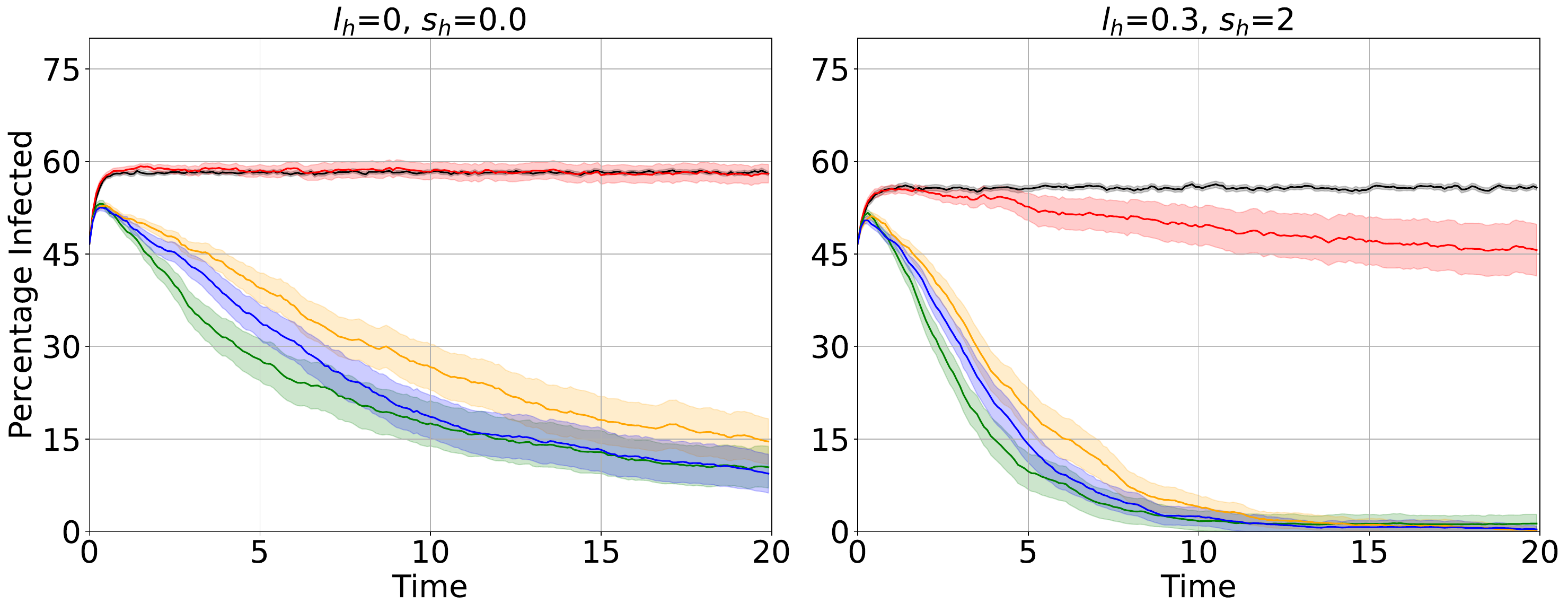}
		\label{fig:smallW_noHD}
    }}
		\vspace{-2em}
	\caption{%
	\textbf{Evolution of nonlinear SIS processes with and without competition, under various intervention strategies.}~%
Percentage of infected nodes under the generalized SIS \Model{eq:genericSIS} using \Eq{eq:spreading_absolute}, when different strategies are employed. Results for ER (a,\,b), PA (c,\,d), and hierER (e,\,f) networks, $300$ nodes in each case. In the plots on the left, healthy states do not diffuse ($\HealFun\op{=}0$). On the right though, there is competition between the positive and negative diffusion ($\HealFun\op{\neq}0$, $\InfFun\op{\neq}0$).
	The model's parameters are set to: $s_{_\InfFun}\op{=}10$, $\slopeSymbol_{_\InfFun}\op{=}2$ for the negative diffusion, and $s_{_\HealFun}\op{=}2$, $\slopeSymbol_{_\HealFun}\op{=}0.25$ for the positive diffusion (when present). At any moment in time, up to $\budget\op{=}10$ nodes are targeted with resource units of $\rho\op{=}155$ healing strength.
	}
\label{fig:competition-lines}
\vspace{-2mm}
\end{figure}

\inlinetitle{Evaluation metrics}{.}
There are various metrics for evaluating the quality of a strategy, \eg expected extinction time, final percentage of infection, area under the curve (AUC). We choose as main measure the AUC as it provides useful measurements even if a strategy does not remove completely the infection, which is a limitation of the expected extinction time. AUC accounts for the total number of infected nodes throughout time, which, in a socioeconomic context, is more interesting than the final infection size. Further, by fixing using the diffusion parametrization $\{s_{_\InfFun},\slopeSymbol_{_\InfFun},s_{_\HealFun},\slopeSymbol_{_\HealFun}\}$, we compare \methodinitials to a competitor by:
\begin{equation}\label{eq:auc-ratio}
\text{AUC}{\text{-ratio}} = \frac{\text{AUC}(\text{\methodinitials})}{\text{AUC}(\text{Competitor})}.
\end{equation}
\subsection{Random Networks}
In this section, we present comparative experiments in random networks. First we introduce gradually nonlinearity and saturation in the diffusion, and then we also show the effects of competition.
To this end, we use \Erdos\ (ER), Preferential Attachment (PA), and Hierarchical \Erdos\ (hierER) random networks of comprising $300$ nodes each. The hierER networks have hierarchical community structure. First, isolated \Erdos\ clusters are created at the lowest level of the hierarchy. Then, a cluster at the $l$-th level is formed by combining a non-intersecting subset of clusters of the $l\!-\!1$-th level, by adding \Erdos\ inter-cluster connections that become weaker as $l$ grows.

\inlinetitle{Introducing nonlinearity and saturation}{.}
We first consider only the negative diffusion (\ie $\HealFun=0$) and we gradually increase $\slopeSymbol_{_\InfFun}$ in an ER random graph, moving gradually from linear (as in the standard SIS model) to nonlinear saturating functions.
\Fig{fig:seq} shows the average over $1000$ simulations of the percentage of infected nodes over time and
the $95$\% confidence interval under the hypothesis of Gaussian distribution of the results.
The results indicate that in the presence of a saturating infection rate, our strategy becomes much more efficient than the competitors.
\inlinetitle{Introducing competition}{.} %
\Fig{fig:competition-lines} presents the effects of positive diffusion that is embedded in the function $\HealFun$, on ER, PR, and SW random networks. To observe more clearly the role of the strategies, we chose to have a less diffusive healthy state compared to the infected state.
The simulations show that, unlike \methodinitials, the methods of the literature lack modeling power to deal efficiently with this complex setting involving saturation as a result of nonlinearity and competition. On the other hand, \methodinitials does not take into account high-level network structure, which is a strength of the MCM strategy (see the hierER example in \Fig{fig:competition-lines}).%

\begin{figure*}[t]
\footnotesize
	\centering
	\!\!\!\!\figlabel{5pt}{\scriptsize\myfont \emph{Polblogs}\phantom{)}\!\!} \phantom{----------------} \figlabel{5pt}{\scriptsize\myfont \emph{{Scenario without competition}} ($\HealFun = 0$)} \phantom{----------------} \figlabel{5pt}{\scriptsize\myfont \emph{Ugandan health advice}\phantom{)}\!\!}\\
	\makebox[\linewidth][c]{%
		\begin{picture}(0,40)
			\put(2,15){\rotatebox{90}{\scalebox{1.}{\tiny {\myfont slope at the origin} ($\slopeSymbol$}\scalebox{.6}{$_{_\InfFun}$}\scalebox{1.}{\tiny)}}}
		\end{picture}
		\hspace{-0.4mm}
	\subfigure[\methodinitials vs LRIE]{
		\includegraphics[width=0.145\linewidth]{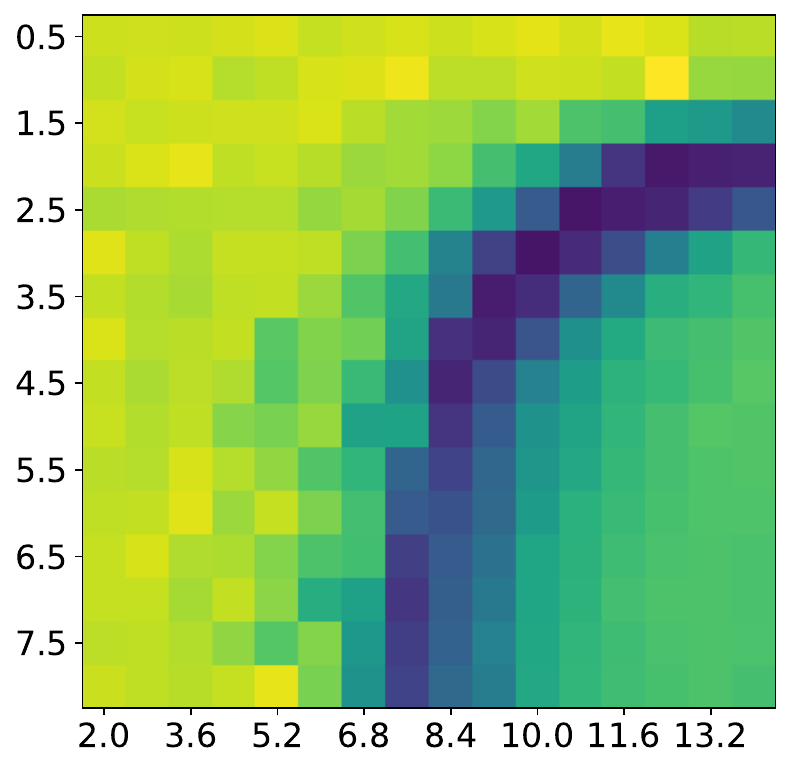}
	    \begin{picture}(0,0)
				\put(-51,-1.5){\scalebox{1.}{\tiny {\myfont saturation} ($s$}\scalebox{.6}{$_{_\InfFun}$}\scalebox{1.}{\tiny)}}
			\end{picture}
  \label{sub:heatmap LRIE Polblogs noHD}
	}%
	\hspace{-1.7mm}%
	\subfigure[\methodinitials vs MCM]{
		\includegraphics[width=0.18\linewidth]{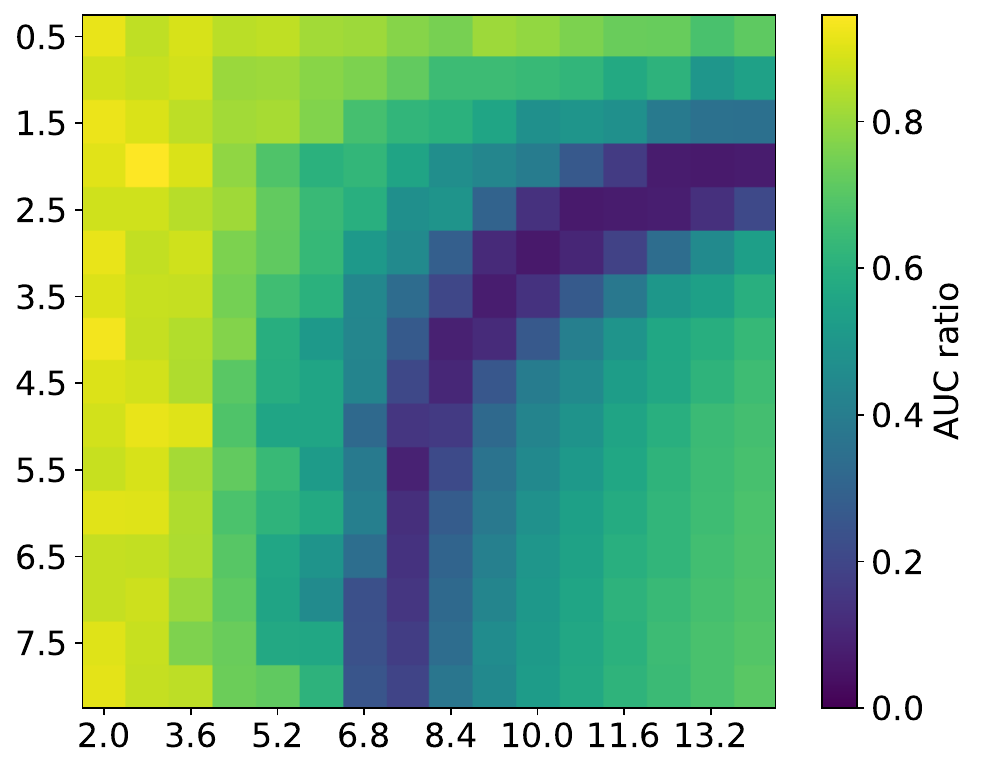}
	    \begin{picture}(0,0)
				\put(-67,-1.5){\scalebox{1.}{\tiny {\myfont saturation} ($s$}\scalebox{.6}{$_{_\InfFun}$}\scalebox{1.}{\tiny)}}
			\end{picture}
		\label{sub:heatmap MCM Polblogs noHD}
	}%
  \hspace{-4mm}%
	\subfigure[\methodinitials: final inf. size]{
		\includegraphics[width=0.183\linewidth]{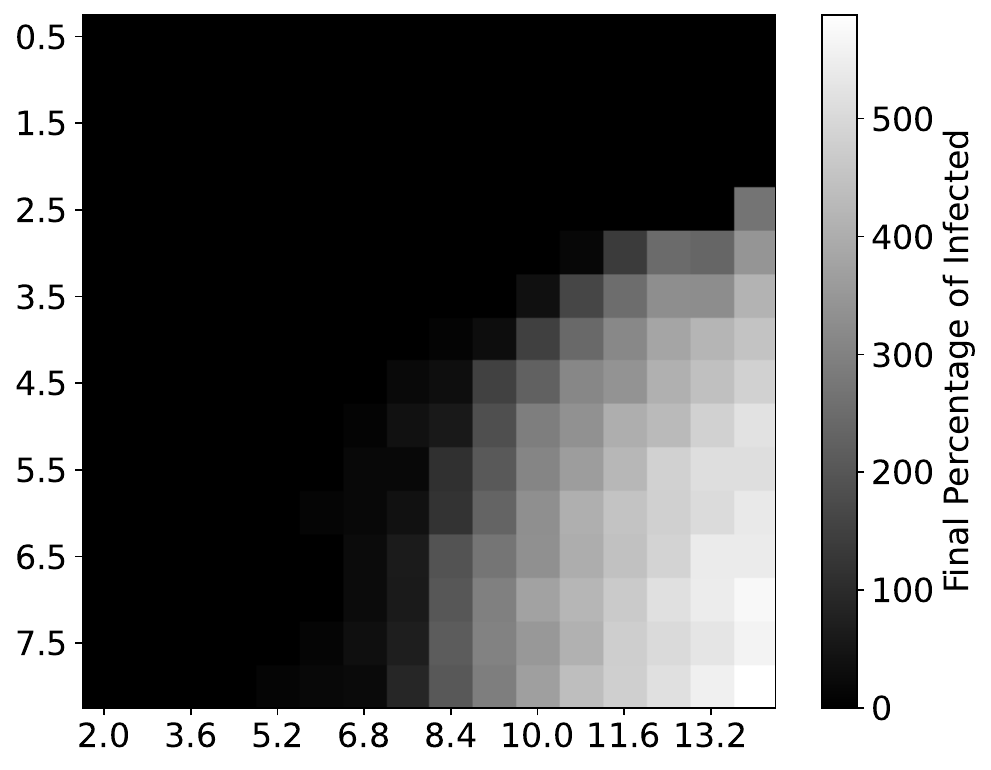}
	    \begin{picture}(0,0)
				\put(-67,-1.5){\scalebox{1.}{\tiny {\myfont saturation} ($s$}\scalebox{.6}{$_{_\InfFun}$}\scalebox{1.}{\tiny)}}
			\end{picture}
		\label{sub:heatmap Polblogs conv noHD}	}%
\hspace{0.em}\tikz{\draw[-,black, densely dashed, thick](0,-1.55) -- (0,1.05);}
  \hspace{0.em}
	\subfigure[gLRIE vs LRIE]{
		\includegraphics[width=0.145\linewidth]{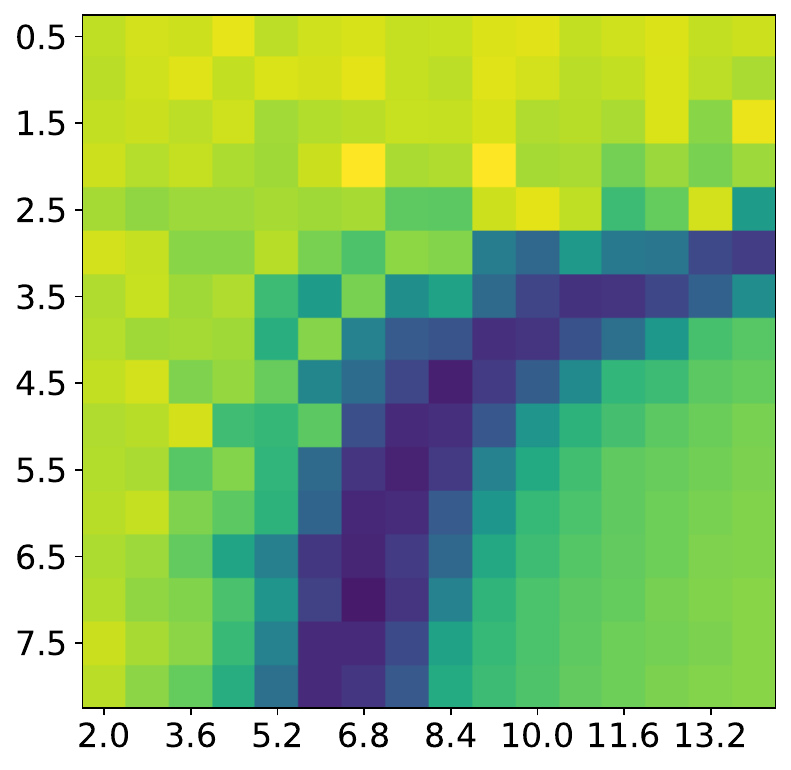}
	    \begin{picture}(0,0)
				\put(-51,-1.5){\scalebox{1.}{\tiny {\myfont saturation} ($s$}\scalebox{.6}{$_{_\InfFun}$}\scalebox{1.}{\tiny)}}
			\end{picture}
	\label{sub:heatmap LRIE health-advice noHD}}%
	\hspace{0.em}%
	\subfigure[gLRIE vs MCM]{
		\includegraphics[width=0.18\linewidth]{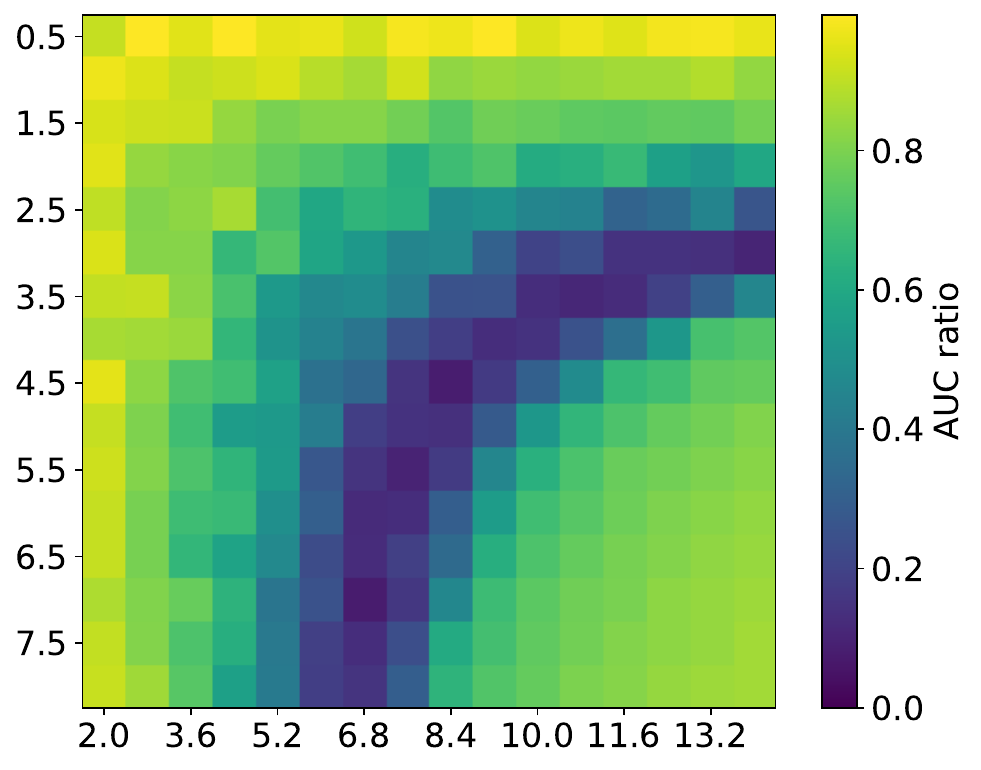}
	    \begin{picture}(0,0)
				\put(-67,-1.5){\scalebox{1.}{\tiny {\myfont saturation} ($s$}\scalebox{.6}{$_{_\InfFun}$}\scalebox{1.}{\tiny)}}
			\end{picture}
		\label{sub:heatmap MCM health-advice noHD}
	}
	\hspace{-4mm}%
	\subfigure[gLRIE: final inf. size]{
		\includegraphics[width=0.183\linewidth]{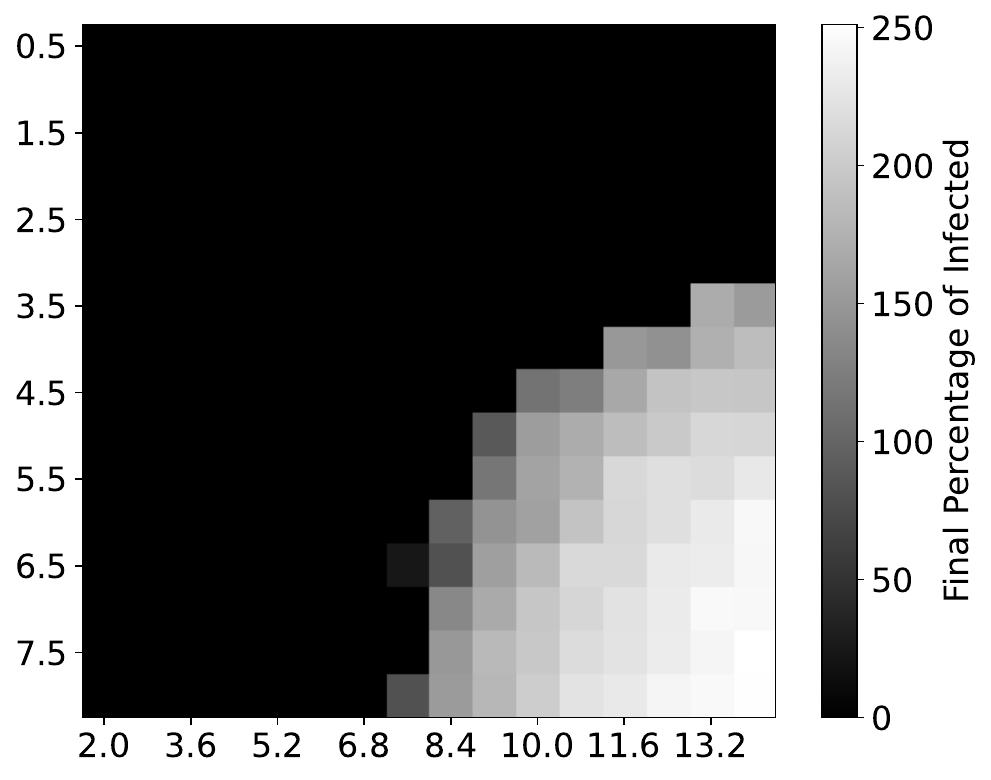}
	    \begin{picture}(0,0)
				\put(-67,-1.5){\scalebox{1.}{\tiny {\myfont saturation} ($s$}\scalebox{.6}{$_{_\InfFun}$}\scalebox{1.}{\tiny)}}
			\end{picture}
		\label{sub:heatmap health-advice conv noHD}
	}}
		\\
		\!\!\!\!\figlabel{5pt}{\scriptsize\myfont \emph{Polblogs}\phantom{)}\!\!} \phantom{-------------------}\figlabel{5pt}{\scriptsize\myfont \emph{{Scenario with competition}} ($\HealFun\neq0$)} \phantom{-------------------} \figlabel{5pt}{\scriptsize\myfont \emph{Ugandan health advice}\phantom{)}\!\!}\\
	\makebox[\linewidth][c]{%
		\begin{picture}(0,40)
			\put(2,15){\rotatebox{90}{\scalebox{1.}{\tiny {\myfont slope at the origin} ($\slopeSymbol$}\scalebox{.6}{$_{_\InfFun}$}\scalebox{1.}{\tiny)}}}
		\end{picture}
		\hspace{-0.4mm}
	\subfigure[\methodinitials vs LRIE]{
		\includegraphics[width=0.145\linewidth]{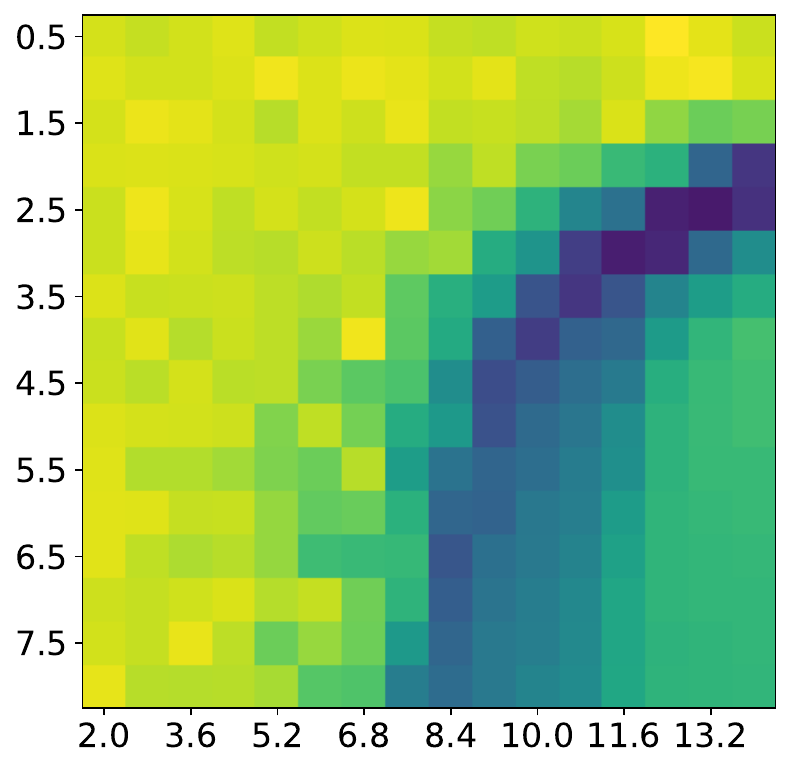}
	    \begin{picture}(0,0)
				\put(-51,-1.5){\scalebox{1.}{\tiny {\myfont saturation} ($s$}\scalebox{.6}{$_{_\InfFun}$}\scalebox{1.}{\tiny)}}
			\end{picture}
  \label{sub:heatmap LRIE Polblogs HD}
	}%
	\hspace{-1.5mm}%
	\subfigure[\methodinitials vs MCM]{
		\includegraphics[width=0.18\linewidth]{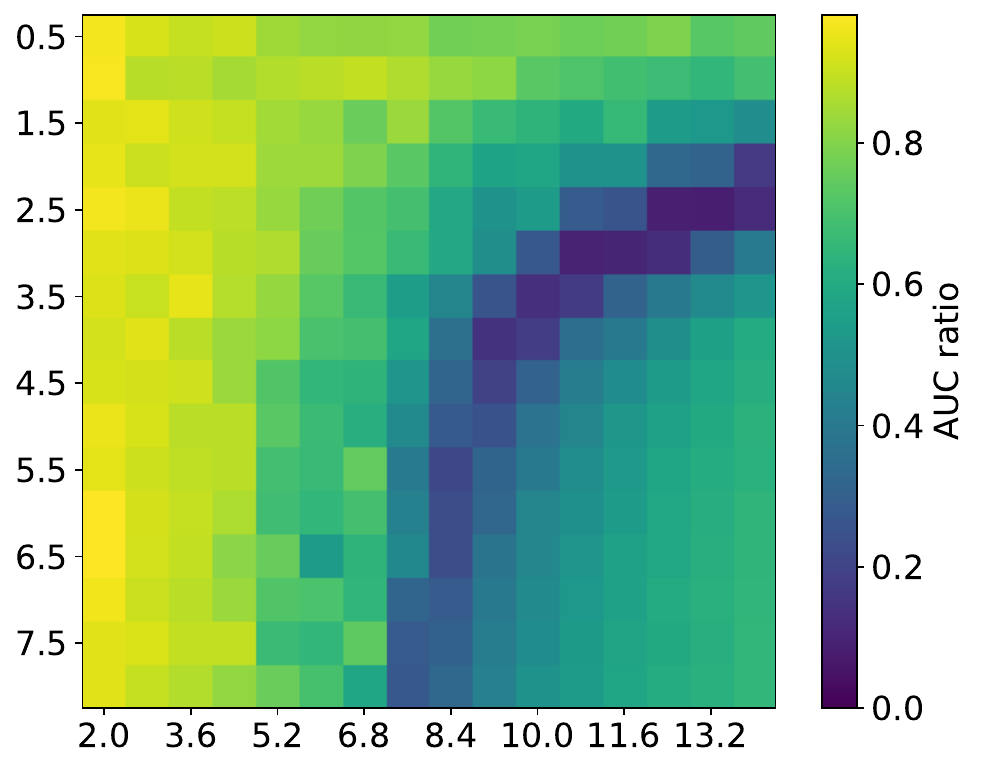}
		\begin{picture}(0,0)
			\put(-67,-1.5){\scalebox{1.}{\tiny {\myfont saturation} ($s$}\scalebox{.6}{$_{_\InfFun}$}\scalebox{1.}{\tiny)}}
		\end{picture}
		\label{sub:heatmap MCM Polblogs HD}
	}%
  \hspace{-4mm}%
	\subfigure[\methodinitials: final inf. size]{
		\includegraphics[width=0.18\linewidth]{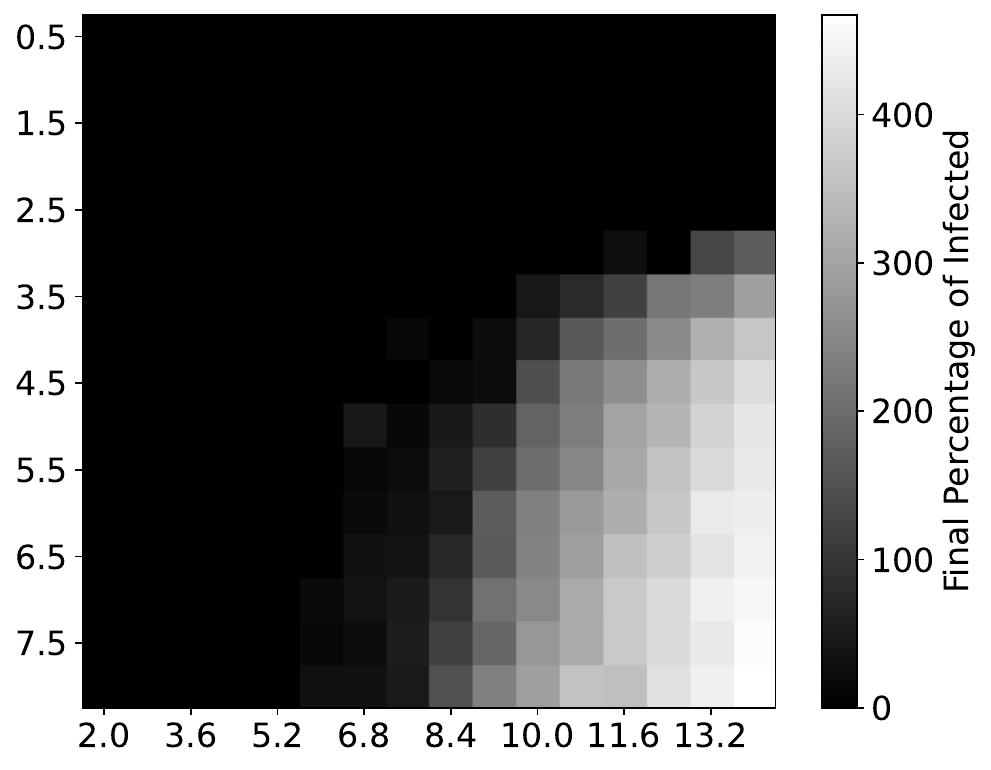}
	    \begin{picture}(0,0)
				\put(-67,-1.5){\scalebox{1.}{\tiny {\myfont saturation} ($s$}\scalebox{.6}{$_{_\InfFun}$}\scalebox{1.}{\tiny)}}
			\end{picture}
		\label{sub:heatmap Polblogs conv HD}
	}
	\hspace{0.em}\tikz{\draw[-,black, densely dashed, thick](0,-1.55) -- (0,1.05);}
	\hspace{0.em}
		\subfigure[gLRIE vs LRIE]{%
			\includegraphics[width=0.145\linewidth]{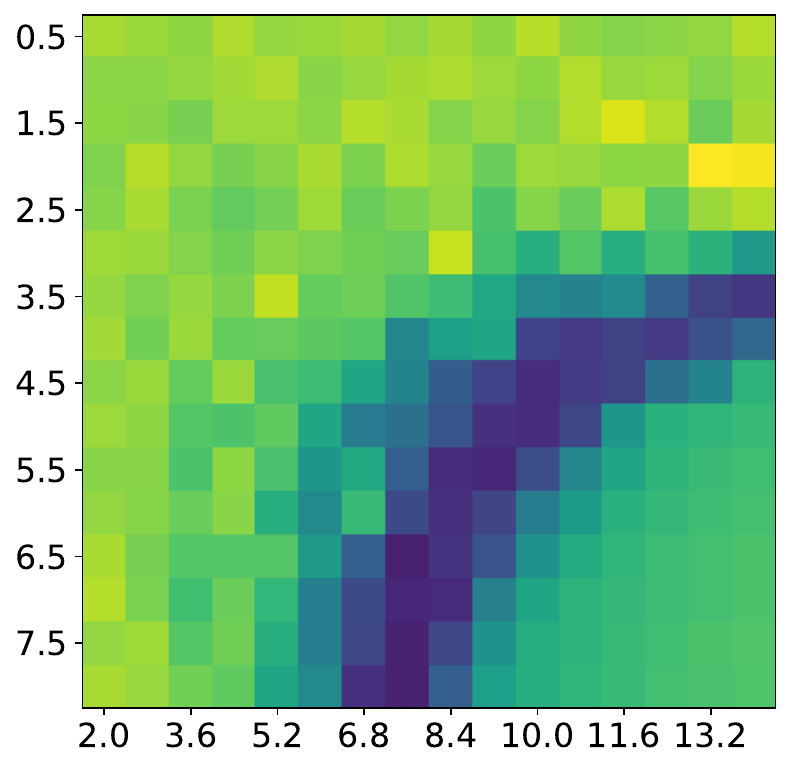}
	    \begin{picture}(0,0)
				\put(-51,-1.5){\scalebox{1.}{\tiny {\myfont saturation} ($s$}\scalebox{.6}{$_{_\InfFun}$}\scalebox{1.}{\tiny)}}
			\end{picture}
	  \label{sub:heatmap LRIE health-advice HD}
		}%
		\!\subfigure[gLRIE vs MCM]{
			\includegraphics[width=0.18\linewidth]{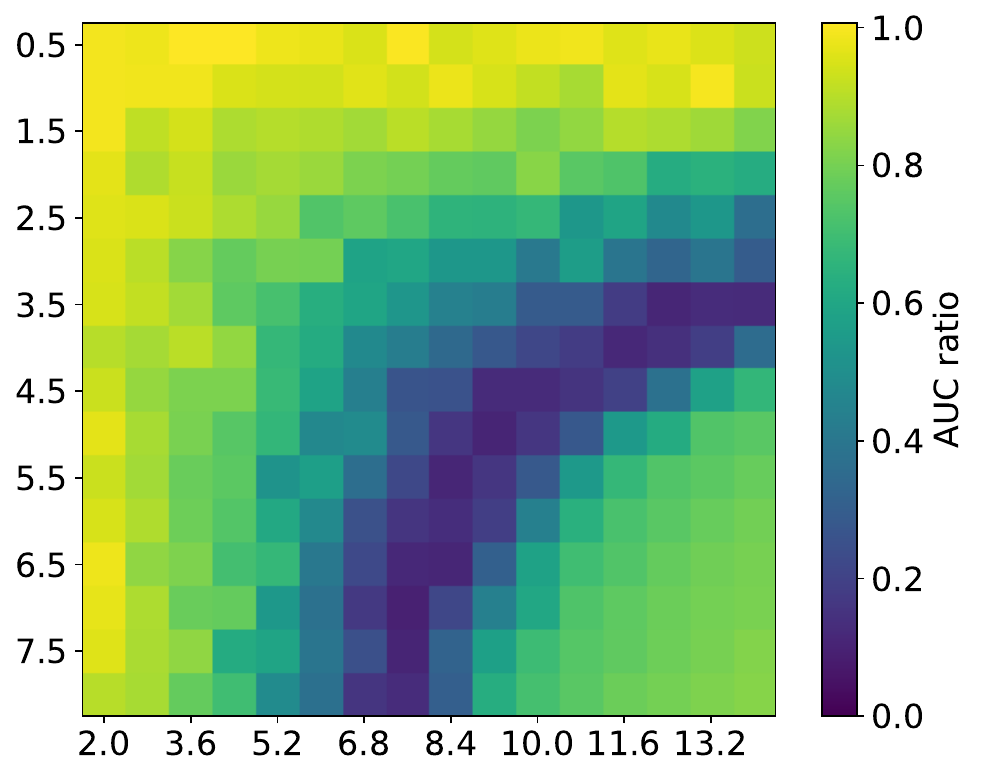}
			\begin{picture}(0,0)
				\put(-67,-1.5){\scalebox{1.}{\tiny {\myfont saturation} ($s$}\scalebox{.6}{$_{_\InfFun}$}\scalebox{1.}{\tiny)}}
			\end{picture}
			\label{sub:heatmap MCM health-advice HD}
		}
		\hspace{-4mm}%
		\subfigure[gLRIE: final inf. size]{
			\includegraphics[width=0.18\linewidth]{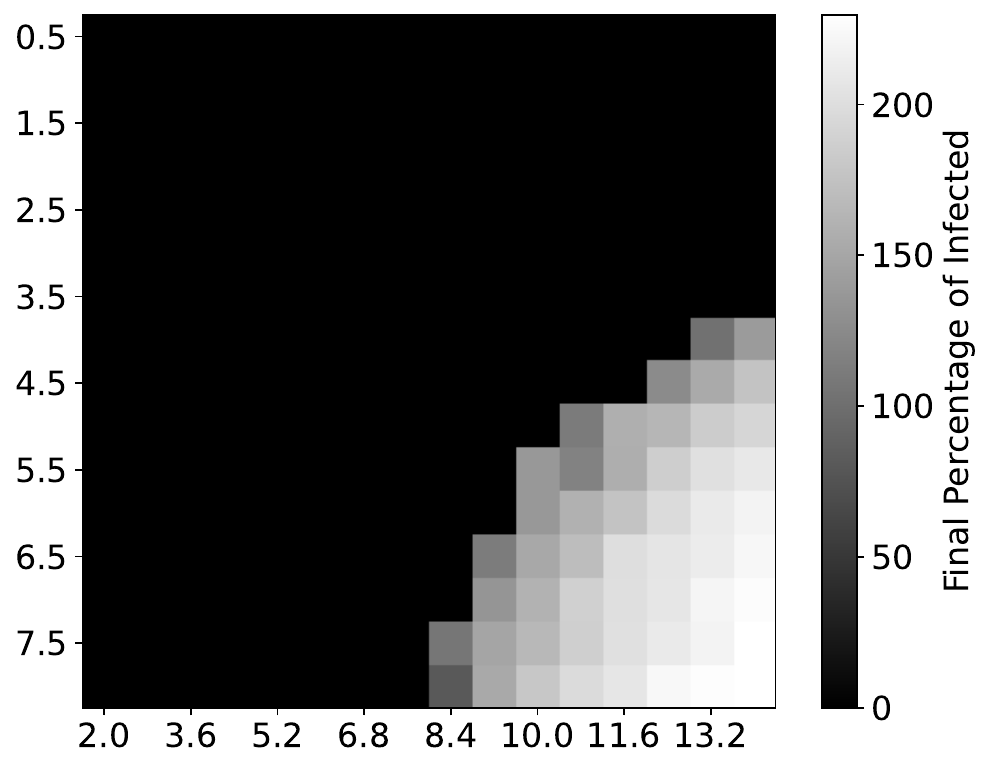}
	    \begin{picture}(0,0)
				\put(-67,-1.5){\scalebox{1.}{\tiny {\myfont saturation} ($s$}\scalebox{.6}{$_{_\InfFun}$}\scalebox{1.}{\tiny)}}
			\end{picture}
			\label{sub:heatmap health-advice conv HD}
		}}
		\caption{\footnotesize\textbf{Pairwise performance comparison between the \methodinitials, LRIE, and MCM intervention strategies via simulations on real datasets.}~Left (resp. right) columns: heatmaps of the AUC-ratio between \methodinitials and competitors, using the generalized SIS model (\Eq{eq:spreading_absolute}) in the Polblogs (resp. health-advice) network with $b=30$ and $\rho = 130$ ($b=10$ and $\rho = 80$). The first row corresponds to the scenario with no positive diffusion, while the second row is the case where positive diffusion is present. The color map ranges from yellow (ratio\op{=}$1$), where the strategies perform the same, to blue (ratio\op{=}$0$), where only \methodinitials manages to remove the infection.
		As expected, the positive diffusion reduces the difficulty of the control problem.
		}
		\label{fig:heat-maps}
	\vspace{-2mm}
\end{figure*}

\subsection{Real Networks}
We performed simulations on the \textit{Polblogs} dataset \cite{adamicPoliticalBlogosphere20042005}, a directed network of $1222$ nodes and $19090$ edges representing hyperlinks between political blogs around the time of the 2004 US presidential election, as well as on the \textit{Ugandan health advice} social network of Ugandan households exchanging health advice in villages \cite{chamiSocialNetworkFragmentation2017}, containing $361$ nodes and $1360$ edges.
Two scenarios were considered, with and without positive diffusion, over a wide parameterization range that allows us to observe the capacity of a strategy to control an infection by affecting a minimal set of agents. %

\Fig{fig:heat-maps} summarizes the empirical results: the first row of heatmaps refers to the scenario with neutralized positive diffusion ($\HealFun=0$), contrary to the second row.
Each grayscale heatmap shows the final infection size when applying \methodinitials; in the black regions the infection is completely removed, while the brighter part indicates persistence at the end of the simulation.
Each colored heatmap, compares \methodinitials to one of the competitors (LRIE and MCM) by showing a landscape of AUC-ratio values (\Eq{eq:auc-ratio}) over a $2$d parametrization. In each heatmap, the shape of the function $\HealFun$ of the positive diffusion is fixed, while the shape of the function $\InfFun$ of the negative diffusion varies: its saturation level increases ($s_\InfFun$) along the x-axis, and its slope ($\ell_\InfFun$) increases along the y-axis. For simplicity, we set for the exogenous rates $\alpha, \delta = 0$, as they merely cause a time shift to the diffusion outcome.

The color heatmaps highlight three regimes. On the top-left side of a heatmap, the difussion parameters define a weak infection for which any strategy would perform well. Contrary, on the bottom-right side, the infection becomes harder to control for all strategies, with the given budget of resources. Moreover, in the left border, the low saturation level causes $\InfFun$ to already saturate with just one infected neighbor. When $\InfFun$ is almost linear and $\HealFun=0$, \methodinitials and LRIE appear to be equivalent.
Overall, those heatmaps exhibit a clear pattern with a right-curve-shaped darker area where \methodinitials overpowers systematically all competitors, showing that it is the most versatile and best performing strategy in the considered competitive spreading setting.

\begin{figure*}[t] \footnotesize
  \centering
  \includegraphics[width=0.8\textwidth]{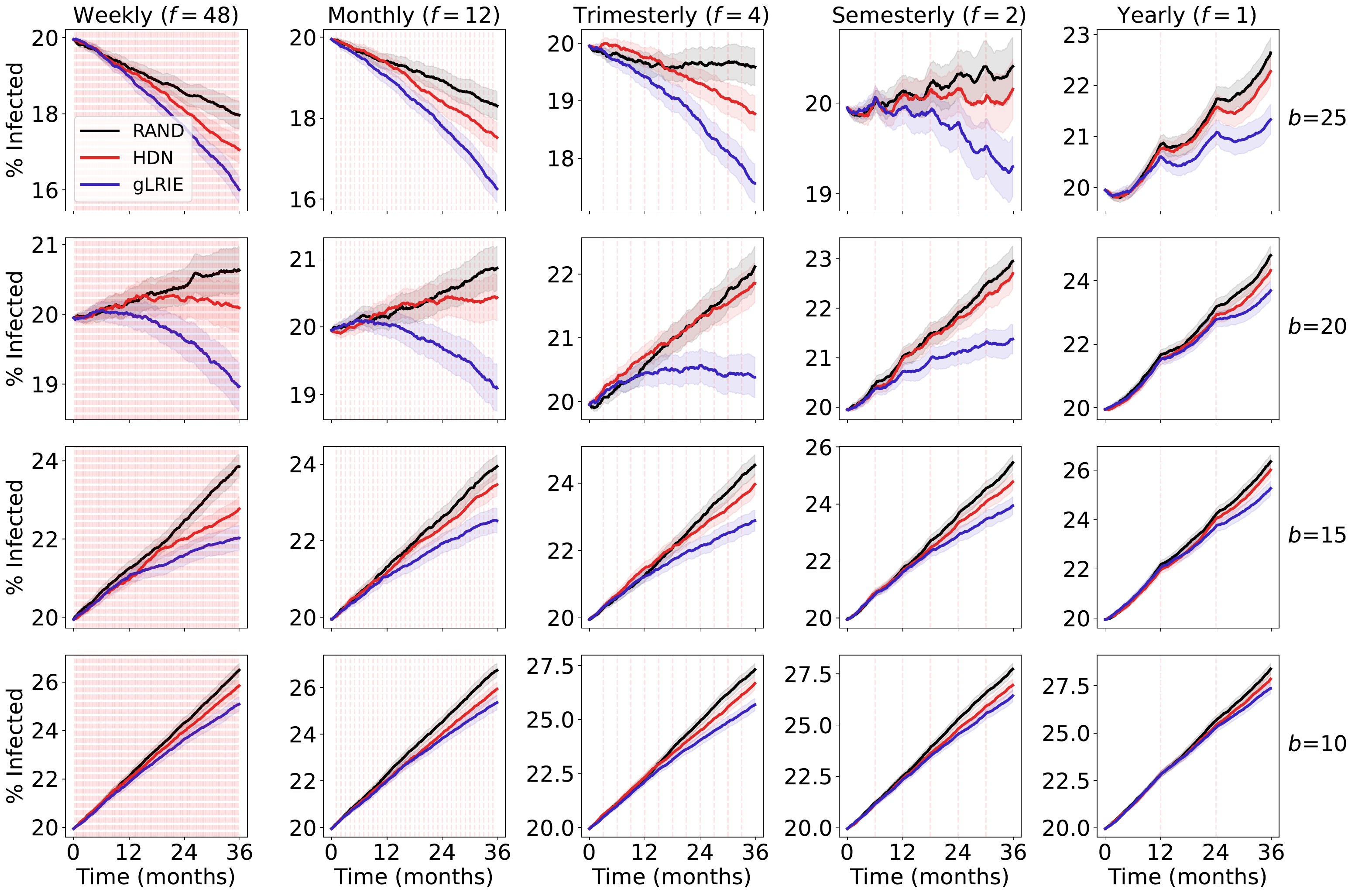}

	\caption{\textbf{Temporal evolution of the percentage of vaping students in a high school social network over a $3$-year period.}~%
	The generalized SIS model (\Eq{eq:spreading_absolute}) is used with $s_{_\InfFun}\op{=}8$, $\slopeSymbol_{_\InfFun}\op{=}1$ for the negative diffusion, and $s_{_\HealFun}\op{=}8$, $\slopeSymbol_{_\HealFun}\op{=}0.5$ for the positive diffusion. At each intervention time (dashed vertical line), up to $b$ nodes are targeted with resource units of $\rho=40$ healing strength. Averaged results over $400$ simulations are plotted.}
  \label{fig:percentInfVape}
\end{figure*}

\subsection{Semi-synthetic high school anti-vaping campaign}
The final experimental scenario simulates an intervention strategy aiming at reducing the number of vaping students in a monitored high school environment. In response to the emergence of applied intervention programes to prevent nicotine use in the past two decades \cite{campbellInformalSchoolbasedPeerled2008,smoking-ces-DP,hollingworth2012reducing}, research in developing effective strategies to prevent cigarette use have been developed \cite{verma2020optimal,labzai2018optimal}. More recently, a growing body of literature on vaping among adolescents \cite{valenteSocialNetworkInfluences2023,wymanInfluenceVapingPeer2021,habibApplyingSocialNetwork2024} has highlighted the need for effective intervention strategies to curb its spread, \eg through counseling sessions or educational programs. %
Motivated by this problem, we design a series of intervention simulations based on real-world social network data to evaluate the performance of \methodinitials and other strategies.
It should be noted that some of the real campaigns target nodes regardless their state, aiming to increase the self-resistance of susceptible nodes and their positive influence to peers. This can be a future extension, however, we adapt the simulations as
our current model targets infected nodes only.

\inlinetitle{Calibration of the vaping experiment}{.}~We use a publicly available adolescent network from a high school in Spain \cite{ruiz-garciaTriadicInfluenceProxy2023}, which contains $409$ nodes and $8557$ directed edges representing 
static asymmetric influence relationships. This is a class of students considered as a small `closed' community over a $3$-year period, from the $10$th to the $12$th grade, during which the vaping behavior is monitored. 
To ensure realistic dynamics, we calibrate the diffusion model underlying \methodinitials using data from the \emph{Monitoring the Future} study \cite{miechMonitoringFutureNational2024}. This longitudinal survey reports that the percentage of vaping $10$th grade students (with or without nicotine) is approximately $20\%$, which increases to $28\%$ by the $12$th grade in the absence of control. To simulate this background tendency in absence of control, we set the diffusion parameters to $s_{_\InfFun}\op{=}8$, $\slopeSymbol_{_\InfFun}\op{=}2$, $s_{_\HealFun}\op{=}8$, and $\slopeSymbol_{_\HealFun}\op{=}0.5$. The asymmetric diffusion intensity reflects the observation that vaping initiation tends to be more contagious than cessation in adolescent populations.

\begin{figure}
  \centering
  \includegraphics[width=0.8\linewidth]{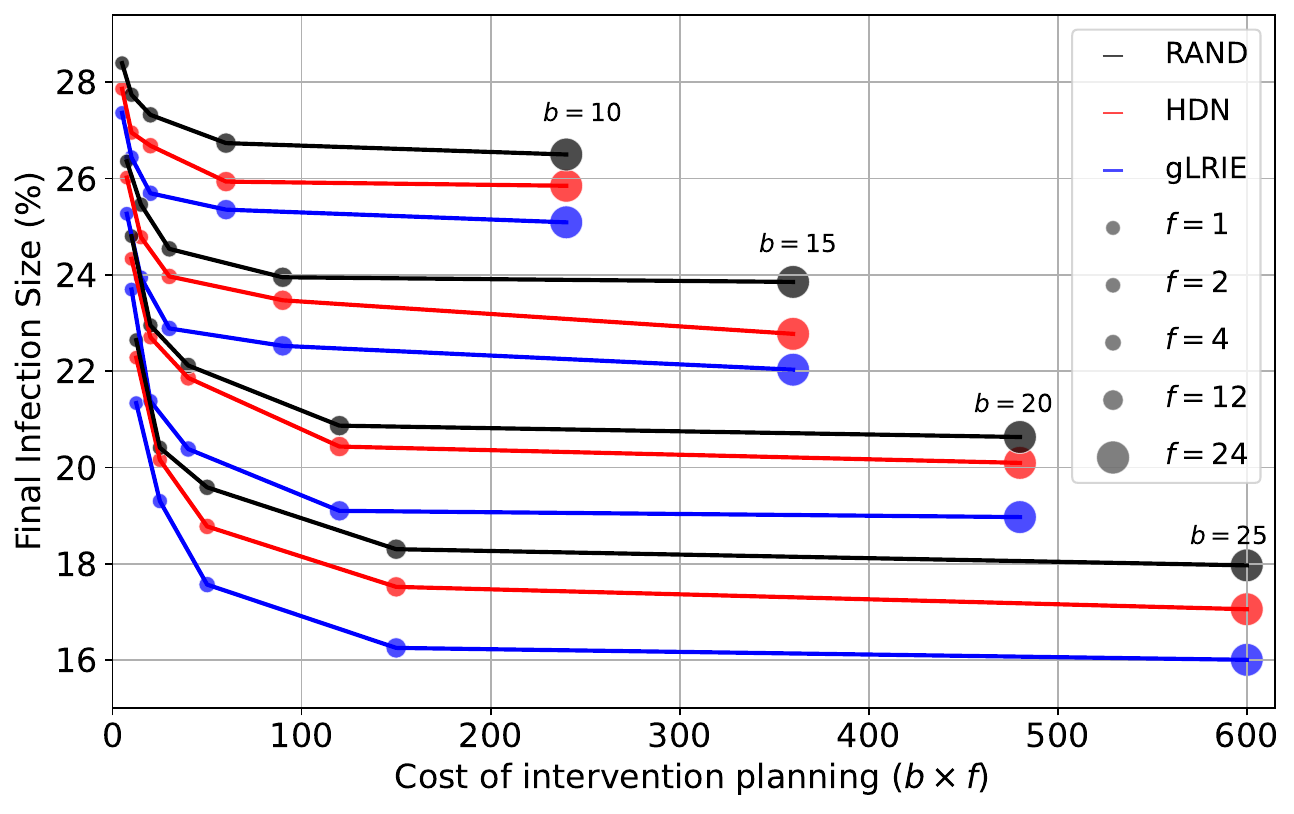}
  \caption{\textbf{Final percentage of vaping students, after a $3$-year period, as a function of total treated students for the RAND, HDN and \methodinitials strategies.
}~Each line corresponds to a different intervention budget ($b$) with varying intervention frequency ($f$), represented by dot sizes.}
  \label{fig:finalInfVapeVsCost}
\end{figure}

\inlinetitle{Intervention protocol}{.}~We evaluate the effectiveness of %
the intervention strategies as a function of three key operational parameters: treatment strength ($\rho$) on an individual, budget ($b$) that is the maximum number of students targeted simultaneously, and the intervention frequency ($f$) corresponding to how often the set of targeted students is revised. To focus on intervention planning, we present results that vary $b$ and $f$, keep fixed $\rho=40$, and simulate over a period of $36$ months, \ie the $3$-year high school duration.
We explore different intervention plans: %
$f=\{12:\text{monthly},\,4:\text{trimesterly},\,2:\text{semesterly},\,1:\text{annually}\}$. %
We compare \methodinitials with two simple and realistic baselines: the random nodes (RAND) and the highest out-degree nodes (HDN) among those being infected. %

\begin{figure}
  \centering
  \includegraphics[width=\linewidth]{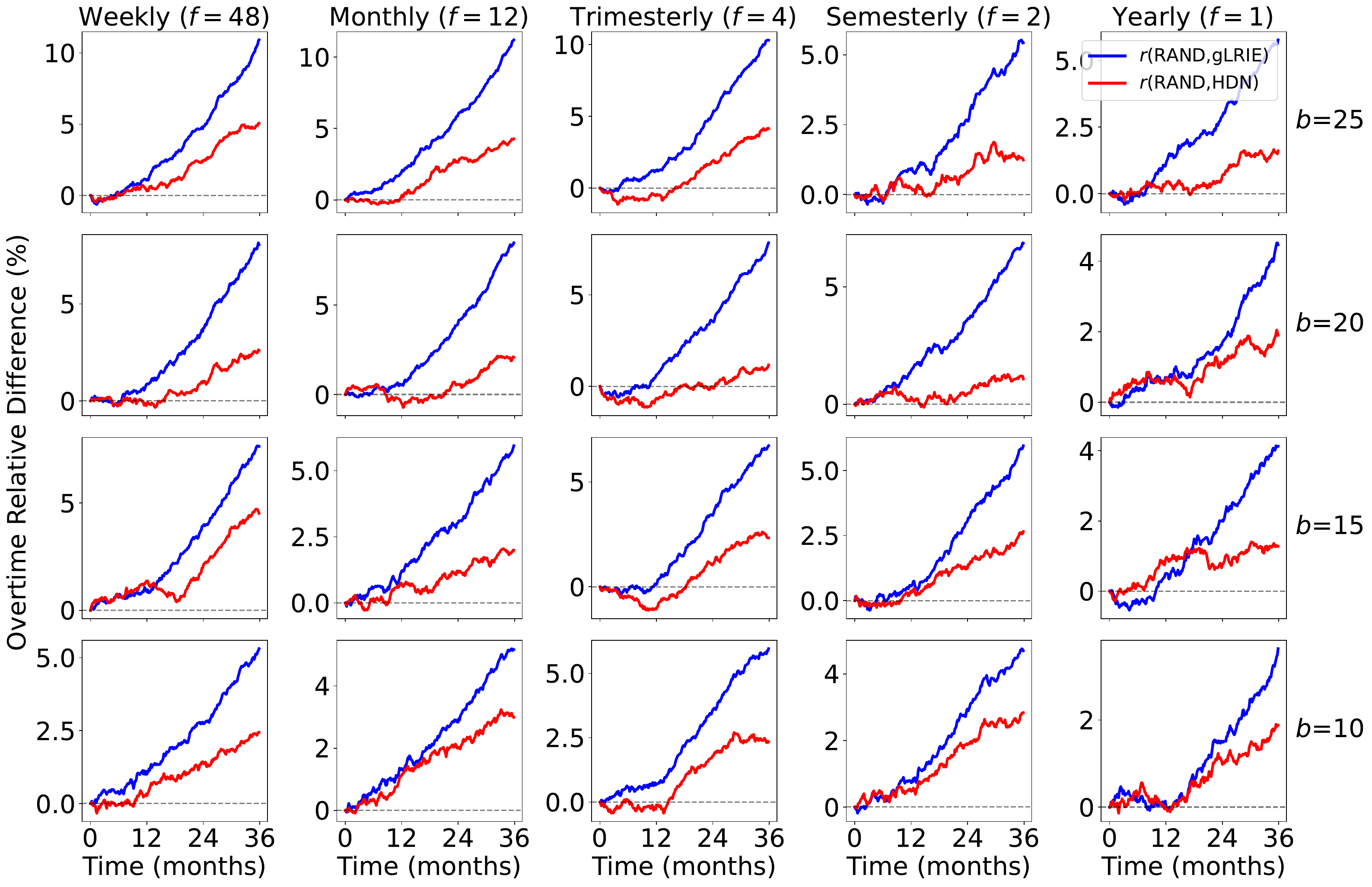}
  \caption{\textbf{Evolution of the relative difference in infection size between the HDN and \methodinitials strategies.}~The effect of each of the compared strategies is expressed relatively to the result of the RAND baseline. The results correspond to the same intervention scenarios as in \Fig{fig:percentInfVape}. Averaged results over $400$ simulations are plotted. Higher positive values imply larger gains compared to RAND.}
  \label{fig:relativeDiffVape}
\end{figure}

\begin{figure*}[t] \footnotesize
  \centering
  \subfigure[$b=10,\,\, f=24$ (weekly)]{
    \includegraphics[width=0.35\textwidth]{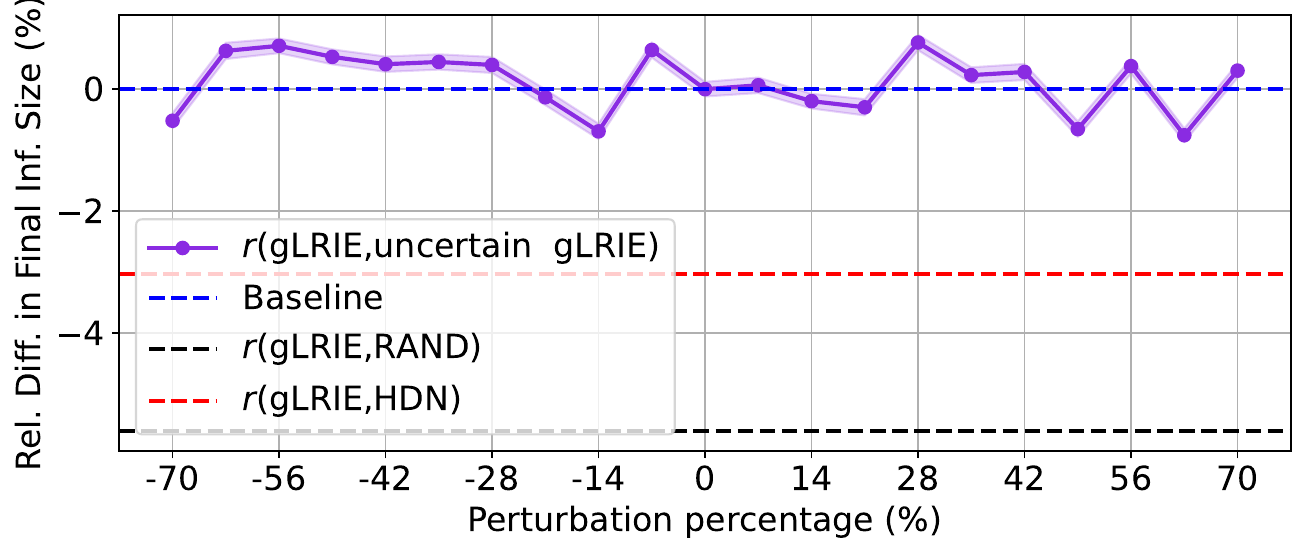}
    \label{subfig:uncertainty_b10}
  }
  \hspace{-1.3em}
  \subfigure[$b=20,\,\, f=2$ (every 6 months)]{
    \includegraphics[width=0.31\textwidth]{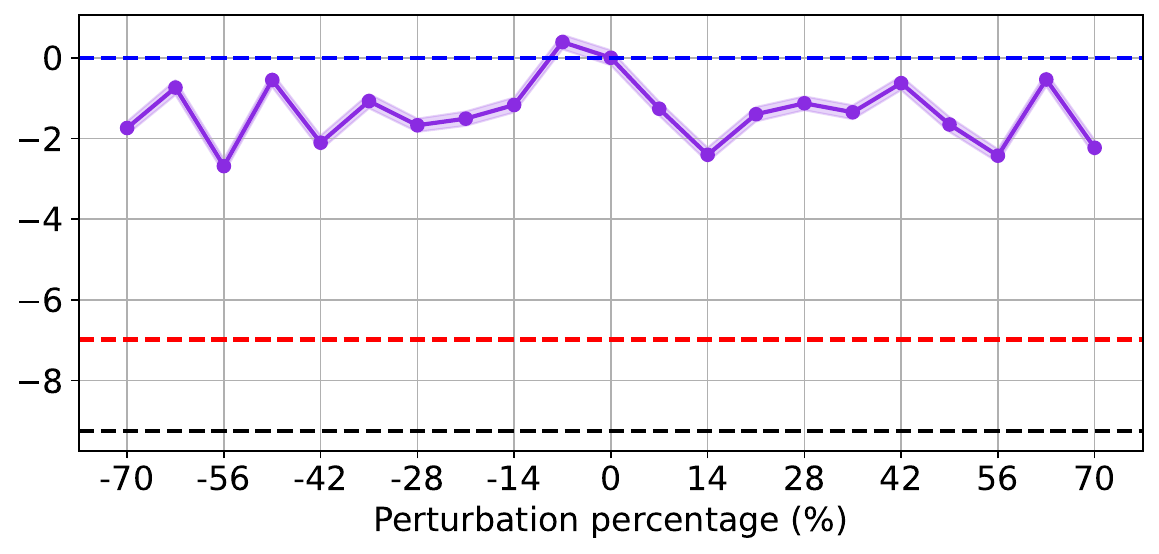}
    \label{subfig:uncertainty_b20}
  }
  \hspace{-1.3em}
  \subfigure[$b=15,\,\, f=1$ (every 12 months)]{
    \includegraphics[width=0.31\textwidth]{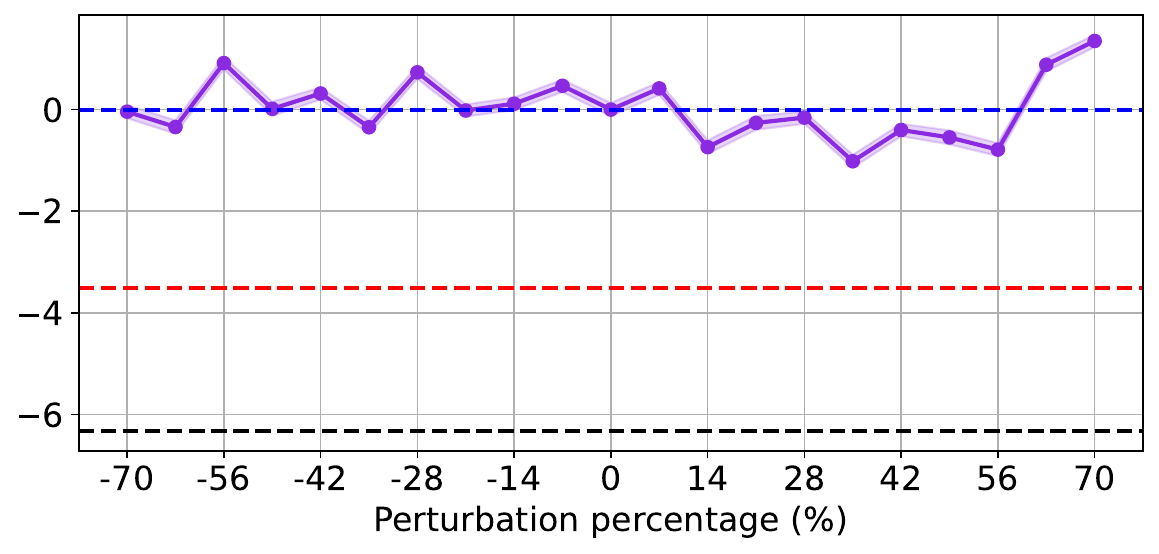}
    \label{subfig:uncertainty_b15}
  }
  \caption{\textbf{Sensitivity analysis of \methodinitials to parameter uncertainty in the high school vaping prevention scenario.}~Final relative difference in final infection size between \methodinitials with the true parameters and \methodinitials with uncertain parameterization is shown for varying uncertainty level considered in the latter case. Dashed lines correspond to the final relative difference between \methodinitials with true parameters and its competitors. The three subfigures present different intervention scenarios in terms of budget ($b$) and frequency ($f$). %
  Averaged results over $400$ simulations are plotted.}
  \label{fig:uncertainty_analysis}
\end{figure*}

\inlinetitle{Results}{.}~\Fig{fig:percentInfVape} illustrates the temporal evolution of the percentage of vaping students under the different intervention strategies. Similarly to the heatmaps presented in \Fig{fig:heat-maps}, the cases at the top left side are easier to deal with, while those at the bottom right are harder.
The results demonstrate that, although the difference is not substantial in every parametrization scenario, \methodinitials is consistently the best strategy over the time period, followed by HDN and RAND. Notably, at high budget, a decrease of the vaping prevalence is visible for \methodinitials but not %
for the other strategies. %
Next, we evaluate the relationship between the final percentage of vaping students and the overall campaign cost expressed as $b \times f$.
In \Fig{fig:finalInfVapeVsCost}, lines represent different budget levels, with varying frequencies indicated by dot sizes. In addition to \methodinitials's superior performance, %
the results highlight that budget has greater contribution than frequency in reducing vaping prevalence. This suggests that coordinated interventions targeting more students at once are more effective than allocating resources thinly over time, whose effect tends to saturate faster.
In the simulated time-frame, the absolute difference in infection percentage may seem moderate at a first glance. However, \Fig{fig:relativeDiffVape} shows the relative difference over time in infection size between HDN and \methodinitials, when the effect of each of them is expressed relatively to the result of RAND. This allows to better appreciate the consistent advantage of \methodinitials, which, in strong intervention scenarios, achieves more than $10\%$ relative reduction in vaping prevalence compared to RAND, whereas HDN only reaches $5\%$. It can be observed that the effects of \methodinitials (resp. HDN) start deviating at around $6$ months (resp. $1$ year) in the simulation for most scenarios, which suggests that early interventions may not yield immediate benefits, but their impact accumulates over time.

\inlinetitle{Sensitivity and feasibility analysis -- Impact.}~It should be noted that \methodinitials is the only method that requires information %
about both the network structure and the diffusion parameters, which may not always be available in practice. In contrast, HDN relies solely on network topology, making it more feasible to deploy in real-world applications.
It is therefore natural to assess the performance gain of \methodinitials in view of its requirement for additional information. %
To this end, we conduct a sensitivity analysis to assess the \methodinitials's robustness to inaccuracies (over- or under-estimation) in the used slope and saturation parameters involved in the infection and healing diffusion functions used in the \methodinitials score computation. %
The results in \Fig{fig:uncertainty_analysis}, indicate that \methodinitials maintains a significant performance advantage over HDN across different scenarios, even with substantial parameter misestimation. %

This semi-synthetic use-case provides empirical evidence about the potential of considering intervention strategies that leverage both network information and positive diffusion, to optimize resource allocation in small-scale and well-monitored social environments, such as public health campaigns within school settings, penitentiary or healthcare institutions. Notably, distinct effects are observed: increasing the budget ($b$) leads to a more pronounced reduction in vaping prevalence, as more students can be targeted simultaneously. Conversely, increasing the intervention frequency ($f$) results in a more temporally %
smooth and sustained effect over time, but of rather smaller effect in reducing the magnitude of the contagion. Moreover, we should stress that the impact of the reported gains %
in the 3-year simulated time-frame is better understood if seen in the frame of a longer process. More specifically, reducing vaping in a class that gets closer to graduation means also less influence to the newer classes that get promoted to the high school. This implies a strong accumulation affect, provided the campaign would run for longer period.

Besides, the vaping use-case is an example of outbreak that starts from a rather low infection level. This falls in the regime where securing high degree nodes is critical due to their high virality toward healthy peers, and refined targeting decisions for other nodes are less impactful. This is exactly what the simulation results portray. However, in scenarios where the initial infection size is larger, \methodinitials would have stronger effect as our earlier synthetic experiments suggest. For instance, according to related studies, the residents of nursing homes suffering from depression are reported to be often at percentages over $40$\% (\eg see \cite{chaoEffectsGroupReminiscence2006}).

\section{Conclusion}\label{sec:conclusion}

In this paper we discussed a general form of recurrent two-state continuous-time Markov process that allows nonlinear node transition functions and competition among the two states that can be both diffusive. This scenario is relevant for a variety of social contagion phenomena where two opposed behaviors spread through a social network. We then proposed the \emph{\methodname} (\methodinitials) algorithm which determines a greedy strategy to suppress the diffusion of the undesired state and promote the preferred one.

Synthetic experiments showed that, compared to competitors from related literature, \methodinitials is well-adapted to the considered setting of competitive spreading, and makes better useof the resources available by targeting the infected nodes that are most critical for the reduction of the contagion.
Finally, we illustrated the effectiveness of \methodinitials in a context of a small well-monitored environment, specifically, vaping prevention in a high school social network. Comparing \methodinitials with strategies that are established in real campaigns, we aimed at providing evidence on the importance of tailored interventions that account for more complex diffusion dynamics. Additionally, the multi-parameter exploration offers actionable insights for administrators who must balance intervention effectiveness with resource constraints and practical feasibility.

Future work could study more complex interaction among the epidemic states (\eg non-exclusivity between strains), %
and the estimation of the diffusion functions from real data that would facilitate the application of \methodinitials in practical scenarios.

\bibliographystyle{IEEEtran}
%

%

\subsection*{Acknowledgments}

A.K. and G.A. acknowledge support from the Industrial Data Analytics and Machine Learning Chair hosted at ENS Paris-Saclay, Université Paris-Saclay. S.S.M. is supported by the Wallenberg AI, Autonomous Systems, and Software Program (WASP).%
\appendix
\label{sec:appendix}
\section*{Analytical derivation of the \methodinitials score}
\begin{proof} Equation \ref{eq:genericScore}\\
Let ue begin by considering the expansion, with respect to $u$, of the cost function introduced in \Eq{eq:minfuncA}: %
\begin{IEEEeqnarray*}{rCl}
Cost(\Lambda, X, \gamma)
&=& \int_0^{\infty} e^{-\gamma u}
\Exp[N_I(t+u)|X(t)=X]du \\
&=& \frac{1}{\gamma}\Phi_{t,X}(0)
 + \frac{1}{\gamma^2}\Phi_{t,X}^{'}(0)
 + \frac{1}{\gamma^3}\Phi_{t,X}^{''}(0)
 + O\Big(\frac{1}{\gamma^4}\Big),
\end{IEEEeqnarray*}
where $\Phi_{t,X} = \Exp[N_I(t+u)|X(t)=X]$.

\inlinetitle{First order term}{.}~This is trivial and does not contain any useful information:
\begin{equation*}
	 \Phi_{t,X}(0) = \sum_i X_i.
\end{equation*}

\inlinetitle{Second order term}{.}~The derivative of a generic function $\mathcal{F}$ is defined as:
\begin{equation*}
	\frac{\partial}{\partial t} \Exp[\mathcal{F}(t)] = \lim_{\Delta t \to 0} \frac{ \Exp[\mathcal{F}(t+\Delta t)]- \Exp[\mathcal{F}(t)]}{\Delta t}.
\end{equation*}\\
The second term:
\begin{IEEEeqnarray*}{rCl}
\Phi_{t,X}^{'}(0)
&=& \bigg\{\sum_i \frac{\partial}{\partial u}
 \Exp[X_i(t+u)|X(t)]\bigg\}\bigg|_{u=0} \\
&=&\biggl\{ -\sum_i \Exp[\HealFun_i X_i]
 -\rho \sum_i \Exp[B_i X_i]
 +\sum_i \Exp[\InfFun_i \notX[i]] \biggr\}\bigg|_{u=0} \\
&=& -\sum_i \HealFun_i X_i
 -\rho \sum_i B_i X_i
 +\sum_i \InfFun_i \notX[i],
\end{IEEEeqnarray*}
which suggests that the treatments are effective only if applied to infected nodes, since they are curing and not preventive resources.

\inlinetitle{Third order}{.}~This term is a bit more complicated:%
\begin{align*}
	 \Phi_{t,X}^{''}(0) =& \sum_i \frac{\partial^2}{\partial u^2} \Exp[X_i(t+u)|X(t)]|_{u=0}\\
	=&-\sum_i \frac{\partial}{\partial u}\Exp[\HealFun_i X_i]|_{u=0} -\rho \sum_i \frac{\partial}{\partial u}\Exp[B_i X_i]|_{u=0} \\
	&+ \sum_i \frac{\partial}{\partial u}\Exp[\InfFun_i \,\notX[i]]|_{u=0}
\end{align*}
\noindent We will consider all the terms separately. \\
At order $\bigO(\Delta t)$:%
\begin{IEEEeqnarray*}{l}
\Exp[\HealFun_i(t+\Delta t)X_i(t+\Delta t)]\\
= \Exp\Biggl[
 \notX[i]\Delta t \InfFun_i \HealFun_i + X_i (1-\Delta t \rho B_i)
 \biggl\{
  \sum_{j \not=i} X_j (\HealFun_j+\rho b_j)
  \HealFun_i^{-j}\Delta t \\
 \quad + \prod_{j \not=i}
 [X_j(1-(\HealFun_j+\rho b_j)\Delta t)
 + \notX[j](1-\InfFun_j\Delta t)]
 \HealFun_i \\
 \quad + \sum_{j \not=i}
 \notX[j]\InfFun_j \HealFun_i^{+j}\Delta t
 \biggr\} - X_i \Delta t
 \biggl\{
  \sum_{j \not=i} X_j (\HealFun_j+\rho b_j)
  (\HealFun_i^{-j})^2\Delta t \\
 \quad + \prod_{j\not=i} [X_j(1-(\HealFun_j+\rho b_j)\Delta t)
 + \notX[j](1-\InfFun_j\Delta t)]
 (\HealFun_i)^2 \\
 \quad + \sum_{j \not=i}
 \notX[j]\InfFun_j (\HealFun_i^{+j})^2\Delta t
 \biggr\}
\Biggr] + \bigO(\Delta t^2),
\end{IEEEeqnarray*}%
where $\HealFun_i^{+j} = \HealFun_i(X_1,X_2,...,X_j=0,...,X_N)$, $\HealFun_i^{-j} = \HealFun_i(X_1,X_2,...,X_j=1,...,X_N)$, and analogously for the infection function $\InfFun_i^{\pm j}$.
The result is:
\begin{IEEEeqnarray*}{ll}
	\frac{\partial}{\partial u}\Exp[X_i\HealFun_i] =& \Exp\Biggl\{\notX[i]\InfFun_i\HealFun_i-X_i(\rho B_i+\HealFun_i)\HealFun_i \\
	&\qquad+ X_i \biggl[-\sum_{j \not=i}(\HealFun_i-\HealFun_i^{-j})X_j(\HealFun_j+\rho b_j) \\
	&\quad\ \ \quad\qquad+\sum_{j \not=i}(\HealFun_i^{+j}-\HealFun_i) \,\notX[j]\InfFun_j\biggr]\Biggr\}.
\end{IEEEeqnarray*}
Before considering the second term, let us make a note: in a small time variation $\Delta t$ the probability of observing $k$ changes goes as $\bigO(\Delta t^k)$. In particular, the probability of no changes and one change in the number of infected in $\Delta t$ are:
\begin{IEEEeqnarray*}{ll}
	\mathbb{P}[N_I(t+\Delta t)=k|N_I(t)=k] \\
	\quad=\prod_i\bigl[X_i(1-(\rho B_i+\HealFun_i)\Delta t)+\notX[i] (1-\InfFun_i \Delta t)\bigr]+\bigO(\Delta t^2) \\
	\quad= 1-\Delta t \sum_i\bigl[X_i(\rho B_i+\HealFun_i)+\notX[i]\,\InfFun_i]+\bigO(\Delta t^2)\\
	\mathbb{P}[N_I(t+\Delta t)=k+1|N_I(t)=k] \\ 
	\quad=\sum_i\notX[i]\,\InfFun_i\Delta t\prod_{j \ne i}\bigl[X_j(1-(\rho b_j+\HealFun_j^{+i})\Delta t) \\
	 \qquad\qquad\qquad\qquad\quad+\notX[j](1-\InfFun_j^{+i} \Delta t)\bigr]+\bigO(\Delta t^3) \\
	\quad= \sum_i\notX[i]\HealFun_i\Delta t \prod_{j \ne i}[1+\bigO(\Delta t)]+\bigO(\Delta t^2) \\
	\quad= \sum_i \notX[i]\InfFun_i\Delta t + \bigO(\Delta t^2)\\
	\mathbb{P}[N_I(t+\Delta t)=k-1|N_I(t)=k] \\
	\quad=\sum_i X_i(\rho B_i +\HealFun_i)\Delta t\prod_{j \ne i}\bigl[X_j(1-(\rho b_j+\HealFun_j^{-i})\Delta t) \\ 
	\qquad\qquad\qquad\qquad\qquad\qquad\ \ \ +\notX[j](1-\InfFun_j^{-i} \Delta t)\bigr]+\bigO(\Delta t^3) \\
	\quad= \sum_iX_i(\rho B_i +\HealFun_i)\Delta t \prod_{j \ne i}[1+\bigO(\Delta t)]+\bigO(\Delta t^2) \\
	\quad= \sum_i X_i(\rho B_i +\HealFun_i)\Delta t + \bigO(\Delta t^2).
\end{IEEEeqnarray*}
We can now consider the second term:
\begin{align*}
	 \sum_i \Exp[B_i X_i] & = \Exp\bigl[\sum_i B_i X_i\bigr] = \Exp[\min(\budget,N_I(t))].
\end{align*}
Let 
	$H(t) = \min (\budget,N_I(t))$.
Then using the previous observation:
\begin{IEEEeqnarray*}{ll}
	\Exp[H(t+\Delta t)]|H(t)>\budget] = \budget+\bigO(\Delta t^2),
\end{IEEEeqnarray*}	
\begin{IEEEeqnarray*}{ll}
	\Exp[H(t+\Delta t)]|H(t)=\budget] \\
	\quad= \budget\biggl\{1-\Delta t \sum_i\bigl[X_i(\rho B_i+\HealFun_i)+\notX[i]\InfFun_i]\biggr\}\\
	\quad\ \ \ + (\budget-1)\biggl\{\sum_i X_i(\rho B_i +\HealFun_i)\biggr\}\Delta t+\bigO(\Delta t^2) \\
	\quad =  \budget-\sum_i X_i(\rho B_i +\HealFun_i)\Delta t+\bigO(\Delta t^2),
\end{IEEEeqnarray*}	
\vspace{-1em}
\begin{IEEEeqnarray*}{ll}
	\Exp[H(t+\Delta t)]|H(t)<\budget] \\
	\quad = N_I(t)\biggl\{1-\Delta t \sum_i\bigl[X_i(\rho B_i+\HealFun_i)+\notX[i]\InfFun_i]\biggr\}\\
	\quad\ \ \ + (N_I(t)-1)\biggl\{\sum_i X_i(\rho B_i +\HealFun_i)\biggr\}\Delta t \\
	\quad\ \ \ +(N_I(t)+1)\biggl\{\sum_i \notX[i]\InfFun_i\biggr\}\Delta t+\bigO(\Delta t^2) \\
	\quad = N_I(t)-\sum_i X_i(\rho B_i +\HealFun_i)\Delta t+\sum_i \notX[i]\InfFun_i\Delta t+\bigO(\Delta t^2),
\end{IEEEeqnarray*}
and finally:
\begin{IEEEeqnarray*}{ll}
	\frac{\partial}{\partial t}\Exp[H(t)] =& -\Exp\biggl[ \Ind{N_I(t)\le \budget}\sum_i X_i(\rho B_i +\HealFun_i)\biggr] \\
	& + \Exp\biggl[\Ind{N_I(t)<\budget}\sum_i \notX[i]\InfFun_i\biggr],
\end{IEEEeqnarray*}
where recall that $\Ind{\cdot}$ stands for the indicator function. This new term does not bring any new information. It suggests, same as the second order term does, to apply the cure only to the infected nodes. The third term is conceptually similar to the first one, and can be split in:
\begin{equation*}
	\frac{\partial}{\partial t}\Exp[\notX[i]\InfFun_i] = \frac{\partial}{\partial t}\Exp[\InfFun_i] - \frac{\partial}{\partial t}\Exp[ X_i \InfFun_i ],
\end{equation*}
\begin{IEEEeqnarray*}{ll}
	\frac{\partial}{\partial t}\Exp[\InfFun_i] \\
	\quad= \Exp\biggl[ \sum_{j \not=i}\InfFun_i^{+j}\notX[j]\InfFun_j-\InfFun_i \sum_{j \not=i}[\notX[j]\InfFun_j-X_j(\rho b_j+\HealFun_j)]\\
	\qquad\quad\ + \sum_{j \not=i} \InfFun_i^{-j} X_j (\rho b_j+\HealFun_j) \biggr] \\
	\quad= \Exp\biggl[\sum_{j \not=i} (\InfFun_i^{+j}-\InfFun_i) \notX[j]\InfFun_j-\sum_{j \not=i}(\InfFun_i-\InfFun_i^{-j})X_j (\rho b_j +\HealFun_j) \biggr],\\
\end{IEEEeqnarray*}	
\begin{IEEEeqnarray*}{ll}
	\frac{\partial}{\partial t}\Exp[X_i \InfFun_i] \\
	\quad= \Exp\Biggl\{\notX[i]\InfFun_i^2-X_i(\rho B_i+\HealFun_i)\InfFun_i + X_i \biggl[\sum_{j \not=i}(\InfFun_i^{+j}-\InfFun_i)\notX[j]\InfFun_j \\
	\qquad\qquad\qquad\qquad\qquad\qquad- \sum_{j \not=i}(\InfFun_i-\InfFun_i^{-j})X_j(\HealFun_j+\rho b_j)\biggr]\Biggr\}.
\end{IEEEeqnarray*}
We can group in the variable $\Xi (t)$ the terms that do not bring new information:
\begin{IEEEeqnarray*}{ll}
	\Phi_{t,X}^{''}(0) \\
	\quad=\rho \sum_i \bigg\{ X_i B_i(\HealFun_i+\InfFun_i) \\
	\qquad\qquad\ \ \ +\sum_{j\not=i}\Big[X_i(\HealFun_i-\HealFun_i^{-j})-\notX[i](\InfFun_i-\InfFun_i^{-j})\Big]X_j b_j \bigg\} +\Xi (t)\\
	\quad=\rho \bigg\{ \sum_i \Big[ X_i B_i(\HealFun_i+\InfFun_i)\Big]\\
	\qquad\qquad\ \ \ +\sum_{\substack{i,j\\ i\not=j}} \Big[X_i(\HealFun_i-\HealFun_i^{-j})-\notX[i](\InfFun_i-\InfFun_i^{-j})\Big] X_j b_j \bigg\} +\Xi (t)\\
\end{IEEEeqnarray*}	
\vspace{-2em}
\begin{IEEEeqnarray*}{ll}
	\quad=\rho \bigg\{  \sum_i \Big[ X_i B_i(\HealFun_i+\InfFun_i)\Big]\\
	\qquad\qquad\ \ \ +\sum_{\substack{i,j\\ i\not=j}} \Big[X_j(\DeltaH{j}{i})-\notX[j](\DeltaI{j}{i})\Big] X_i B_i \bigg\} +\Xi (t)\\
	\quad=\rho \sum_i \bigg\{ \bigg[ (\HealFun_i+\InfFun_i)\\
	\qquad\qquad\ \ \ \ \ \ +\sum_{j\not=i} \Big[X_j(\DeltaH{j}{i})-\notX[j](\DeltaI{j}{i})\Big]\bigg] X_i B_i \bigg\} +\Xi (t).
\end{IEEEeqnarray*}

Finally, in order to minimize the infection, our analysis suggests to heal only infected nodes and consider the score $S_i$ according to the last term:
\begin{equation*}
	S_i = -\bigg[ (\HealFun_i+\InfFun_i)+\sum_{j\not=i}\Big[X_j(\DeltaH{j}{i})-\notX[j](\DeltaI{j}{i})\Big]\bigg].
\end{equation*}
\end{proof}

\end{document}